\documentclass[12pt,aps,floats,showpacs,amssymb,tightenlines]{revtex4}

\usepackage{color}
\usepackage{graphicx}
\usepackage{amsmath}
\usepackage{amsfonts}
\usepackage{amscd}
\usepackage{epsfig}
\usepackage{amssymb}
\usepackage{tabularx}

\begin{document}

\title{Linear stability of the Linet - Tian solution with positive cosmological constant.}
\author{Reinaldo J. Gleiser} \email{gleiser@fis.uncor.edu}

\affiliation{Instituto de F\'{\i}sica Enrique Gaviola and FAMAF,
Universidad Nacional de C\'ordoba, Ciudad Universitaria, (5000)
C\'ordoba, Argentina}

\begin{abstract}

The Linet - Tian metrics are solutions of the Einstein equations
with a cosmological constant, $\Lambda$, that can be positive or
negative. In the limit of vanishing $\Lambda$ they reduce to a form
of the Levi - Civita metric, and, therefore, they can be considered
as generalizations of the former to include a cosmological constant.
The gravitational instability of both the Levi - Civita metric, and
of the Linet - Tian solution with $\Lambda <0$, was recently
established, and the purpose of this paper is to extend those
results to the case $\Lambda > 0$. A fundamental difference brought
about by a positive cosmological constant, already known in the
literature, is in the structure of the resulting space time.
Associated with each of the two commuting Killing vectors
$\partial_{\phi}$, and $\partial_{z}$, there is a curvature
singularity that has the same characteristics as that associated to
$\partial_{\phi}$ in the Levi - Civita metric, and we show that
there is an isometry relating these singularities that reduces the
effective parameter space of the metrics. In attempting to set up
and solve the linearized perturbation equations we are confronted
with the problem of a gauge ambiguity that leads to the introduction
of a gauge invariant function, $W_1$, that is shown to be also a
{\em master function}, that satisfies a second order ODE, and in
terms of which one can express all the perturbation functions.
Unfortunately, the equation satisfied by $W_1$ contains singular
coefficients, and, although one can show that {\em all} its
solutions are regular, because of the presence of these
singularities one cannot, as in the case of negative $\Lambda$, set
up an associated self adjoint problem that provides a complete set
of solutions for $W_1$. We are thus restricted to solving
numerically the perturbation equations, and using those solutions
for constructing $W_1$, for particular values of the parameters. In
all the cases analyzed we find unstable modes, which strongly
suggests that all the Linet - Tian space times with $\Lambda > 0$
are linearly unstable under gravitational perturbations. The problem
of determining the time evolution of arbitrary initial data in terms
of the $W_1$, or something equivalent, remains open.
\end{abstract}


\maketitle

\section{Introduction}

The Linet - Tian metrics \cite{linet}, \cite{tian}, are static
solutions of the Einstein equations with a cosmological constant,
$\Lambda$, that can be positive or negative, that posses also two
commuting Killing vectors: $\partial_{\phi}$, and $\partial_{z}$ .
They are characterized by two constants: one is $\kappa$, associated
to the singularities of the metrics, and the other is the
cosmological constant $\Lambda$. In the limit of vanishing
cosmological constant they reduce to a form of the Levi - Civita
metric \cite{levi}, and, therefore, they can be considered as
generalizations of the former to include a cosmological constant.
Both the Levi - Civita metric, and the Linet - Tian solution with
negative cosmological constant, have been found to be
gravitationally unstable, as was recently established in
\cite{glei1} and \cite{glei2}. The purpose of this paper is to
extend those results to the case of positive cosmological constant.
A fundamental difference brought about by a positive cosmological
constant, already known in the literature,\cite{wang},
\cite{griffiths}, is in the structure of the resulting space time.
While for both the Levi-Civita and Linet-Tian metric with negative
cosmological constant it is natural to assign to the space time a
cylindrical symmetry, associated to the Killing vector
$\partial_{\phi}$, with $\partial_z$ corresponding to translations
along the axis of cylindrical symmetry, in the case of a {\em
positive} $\Lambda$, one finds that associated with each one of
$\partial_{\phi}$, and $\partial_{z}$, there is a curvature
singularity that has the same characteristics as that associated to
$\partial_{\phi}$ in the Levi - Civita metric \cite{wang}, and,
therefore, as indicated in \cite{griffiths}, there is no natural way
to consider the resulting space time as cylindrically symmetric.
Moreover, we find that, up to all diverging terms, the singularities
associated to $\partial_{\phi}$, and $\partial_{z}$ are independent
of $\Lambda$, and are characterized by the same Kretschmann
invariant as for the Levi-Civita metric. These features are analyzed
in detail in Section II where we show that there is an isometry
relating these singularities, that reduces the effective parameter
space of the metrics.

In Section III we introduce a form of the Linet-Tian metric with a
positive cosmological constant, that aims at simplifying the
analysis of its linear stability, by introducing a new ``radial''
coordinate $y$, with a range $0\leq y \leq1$, with $y=0$ the
singularity associated to $\partial_{\phi}$, and $y=1$ to that
associated to $\partial_{z}$. In Section IV we consider a general
linear perturbation of the metric, but independent of $\phi$, in
agreement with the above mentioned isometry. We include a detailed
analysis of the resulting gauge dependence and ambiguities of the
perturbations, and show that the perturbations separate into two
groups that transform independently under coordinate
transformations. This is also in agreement with the form of the
perturbation equations, that show the same separation into the two
independent groups. The present analysis concentrates on one of
these groups, which we call the ``diagonal'' case, and the other is
left for a separate study. In Section V we consider the diagonal
case, and display the corresponding Einstein equations, which in
this case reduce to a system of three linear, first order ODE for
three functions describing the perturbations. We analyze the
possible independent solutions by obtaining their behaviour either
as $y \to 0$, or $y \to 1$. We show that the system admits three
independent solutions: one is a pure gauge solution that can be
given as an exact solution using the results of Section IV. The
other two can be separated into one that approaches a finite limit
either as $y \to 0$, or $y \to 1$, while the other diverges as
$\ln(y)$, as $y \to 0$, or as $\ln(1-y)$ when $y \to 1$.
Unfortunately, as discussed in the text, because of the already
mentioned gauge ambiguities it contains, the system as such is not
adequate for an analysis of the possible unstable modes of the
perturbations. To solve this problem we introduce in Section VI a
{\em gauge invariant function}, $W_1(y)$, that approaches finite
limits for either $y \to 0$, or $y \to 1$, when the corresponding
gauge independent part of the system approaches similar limits.
Moreover, we show that $W_1(y)$ is also a {\em master function}, in
terms of which we can express all the perturbation functions. The
gauge dependence of these functions can easily be seen in those
expressions. We show that $W_1(y)$ satisfies a {\em second order}
linear ODE. Imposing boundary values on the solutions of this ODE we
look for the possible spectrum of unstable modes. A first approach:
changing to a new variable $r=r(y)$ and function
$\widetilde{W}_1(r)$, that satisfies (one dimensional) Schroedinger
like equation, to determine the spectrum, unfortunately fails,
because the resulting ``potential'' is singular. This singularity
can be traced to the fact that the ODE that $W_1$ satisfies has also
singular coefficients, although one can show that all the solutions
are regular in $0 < y < 1$. In view of these difficulties we
recapitulate in Section VII on the nature of the problem we want to
solve, and give the reasons for considering directly a numerical
analysis of the system of ODE's satisfied by the perturbation
functions. This analysis is carried out in Section VIII, for two
particular choices of parameters, after imposing appropriate
boundary conditions, which are explicitly given, at either $y=0$, or
$y=1$. The numerically computed functions are then used to compute
the gauge invariant function $W_1(y)$, and a ``shooting'' approach
is applied to obtain the two lowest eigenvalues. We find that, in
both cases, the lowest eigenvalue corresponds to an unstable mode,
while the next to lowest eigenvalue corresponds to a stable one. The
special cases $\kappa=0$, and $\kappa=1$ are analyzed in Section IX.
Finally, in Section X we give a brief description of the main
results of the paper, and discuss several issues not covered here,
that will be considered in a separate paper.

\section{Some properties of the Linet - Tian metric with a
positive cosmological constant.}

In the case $\Lambda > 0$ (positive cosmological constant), the
Linet - Tian metric  can be locally written in the form,
\begin{equation}\label{LTeq1}
ds^2 = Q^{2/3} \left(-P^{p_1}dt^2+P^{p_2}
dz^2+P^{p_3}d\phi^2\right)+ d\rho^2
\end{equation}
where:
\begin{eqnarray}\label{LTeq2}
   Q(\rho) & = & \frac{1}{\sqrt{3 \Lambda}}\sin\left(\sqrt{3\Lambda}\rho\right) \nonumber \\
   P(\rho) & = & \frac{2}{\sqrt{3 \Lambda}}\tan\left(\frac{\sqrt{3\Lambda}}{2}\rho\right)
\end{eqnarray}
and the parameters $p_i$ satisfy,
\begin{eqnarray}\label{LTeq4}
   p_1+p_2+p_3 & = &  0  \nonumber \\
  p_1{}^2 +p_2{}^2 +p_3{}^2 & = &  \frac{8}{3}
\end{eqnarray}
They may be parameterized as \cite{thorne},
\begin{eqnarray}\label{LTeq5}
   p_1 & = & -\frac{2(1- 2\kappa -2 \kappa^2)}{3(1+\kappa+\kappa^2)}  \nonumber \\
  p_2 & = & -\frac{2(1+4 \kappa +\kappa^2)}{3(1+\kappa+\kappa^2)}    \\
  p_3 & = & \frac{2(2+2\kappa -\kappa^2)}{3(1+\kappa+\kappa^2)}  \nonumber
\end{eqnarray}
\\

It is clear from (\ref{LTeq1},\ref{LTeq2}) that both
$\partial/\partial \phi$, and $\partial/\partial z$, are Killing
vectors. In a cylindrically symmetric metric one usually identifies
$\partial/\partial \phi$ as a ``rotational'' Killing vector,
assuming for $\phi$ a finite range, say $0 \leq \phi \leq 2 \pi$,
with the ends identified, and identifies $\partial/\partial z$ with
a ``translational'' Killing vector, allowing for $z$ the range
$-\infty < z < +\infty$. This situation appears natural in the case
$\Lambda <0$, as, for instance, in \cite{glei1}. On the other hand,
for $\Lambda > 0$, this identification is far more delicate.
Consider again (\ref{LTeq2}). This may be written in the form,
\begin{eqnarray}\label{LTeq2a}
Q(\rho) & = & \frac{1}{2\sqrt{3 \Lambda}}\sin\left(
\sqrt{3\Lambda}\;\rho/2\right)
\cos\left(\sqrt{3\Lambda}\;\rho/2\right) \nonumber \\
   P(\rho) & = & \frac{2}{\sqrt{3
   \Lambda}}\frac{\sin\left(\sqrt{3\Lambda}\;\rho/2\right)}{\cos\left(\sqrt{3\Lambda}\;\rho/2\right)}
\end{eqnarray}
and, therefore, (after some constant rescalings of $(t,z,\phi)$),
and using (\ref{LTeq5}) explicitly, (\ref{LTeq1}) may be written in
the form,
\begin{eqnarray}
\label{LTeq2b}
  ds^2 &=& -\sin\left(\sqrt{3\Lambda}\;\rho/2\right)^{\frac{2 \kappa(1+\kappa)}{1+\kappa+\kappa^2}}
  \cos\left(\sqrt{3\Lambda}\;\rho/2\right)^{-\frac{2 (\kappa+2)(\kappa-1)}{3(1+\kappa+\kappa^2)}} dt^2 \nonumber \\
   & &+ d\rho^2 \\
   & &+\sin\left(\sqrt{3\Lambda}\;\rho/2\right)^{\frac{-2 \kappa}{1+\kappa+\kappa^2}}
  \cos\left(\sqrt{3\Lambda}\;\rho/2\right)^{\frac{2 (\kappa+2)(2\kappa+1)}{3(1+\kappa+\kappa^2)}} dz^2 \nonumber \\
   && +\sin\left(\sqrt{3\Lambda}\;\rho/2\right)^{\frac{2(1+\kappa)}{1+\kappa+\kappa^2}}
  \cos\left(\sqrt{3\Lambda}\;\rho/2\right)^{\frac{2 (2\kappa+1)(\kappa-1)}{3(1+\kappa+\kappa^2)}} d\phi^2 \nonumber
\end{eqnarray}
But, if we now introduce a new coordinate $r$ as follows,
\begin{equation}
\label{LTeq2c}
  \rho = \frac{\pi}{\sqrt{3 \Lambda}} - r
\end{equation}
and a new parameter $\eta$ such that,
 \begin{equation}\label{LTeq2d}
    \kappa =\frac{1-\eta}{2 \eta+1}
 \end{equation}
we find that (\ref{LTeq2b}) takes the form,
\begin{eqnarray}
\label{LTeq2e}
  ds^2 &=& -\sin\left(\sqrt{3\Lambda}\;r/2\right)^{\frac{2 \eta(1+\eta)}{1+\eta+\eta^2}}
  \cos\left(\sqrt{3\Lambda}\;r/2\right)^{-\frac{2 (\eta+2)(\eta-1)}{3(1+\eta+\eta^2)}} dt^2 \nonumber \\
   & &+ dr^2 \\
    & &+\sin\left(\sqrt{3\Lambda}\;r/2\right)^{\frac{-2 \eta}{1+\eta+\eta^2}}
  \cos\left(\sqrt{3\Lambda}\;r/2\right)^{\frac{2 (\eta+2)(2\eta+1)}{3(1+\eta+\eta^2)}}
  d \phi  \nonumber \\
   && +\sin\left(\sqrt{3\Lambda}\;r/2\right)^{\frac{2(1+\eta)}{1+\eta+\eta^2}}
  \cos\left(\sqrt{3\Lambda}\;r/2\right)^{\frac{2 (2\eta+1)(\eta-1)}{3(1+\eta+\eta^2)}} d z^2
  \nonumber
\end{eqnarray}
which is the same as that of (\ref{LTeq2b}), with the replacement of
$\kappa$ by $\eta$ and an exchange of the roles of $z$ and $\phi$.
In other words, any Linet-Tian metric with positive $\Lambda$, and
parameter $\kappa$ is locally isometric to a Linet - Tian metric
with the same $\Lambda$, and parameter $\eta$ related to $\kappa$
through (\ref{LTeq2d}), but with the roles of $z$ and $\phi$
interchanged. This indicates that there is no intrinsic geometric
difference between the Killing vectors $\partial/\partial z$, and
$\partial/\partial \phi$. In fact both represent ``rotations'' about
a symmetry axis. $\partial/\partial \phi$ corresponds to rotations
about the axis at $\rho=0$, and $\partial/\partial z$ to rotations
about the axis at $\rho = \pi/\sqrt{3 \Lambda}$. We may, on this
account, ``naturally'' assume a finite range for {\em both} $z$ and
$\phi$, with the ends identified, so that the integral curves of
both Killing vectors are generally of finite length. But even with
this assumption this length may diverge as $\rho$ approaches either
$0$ or $\pi/\sqrt{3\Lambda}$. The particular behaviour depends on
$\kappa$. In accordance with (\ref{LTeq2b}), as $\rho \to 0$ we
have,
\begin{eqnarray}
\label{Kill01}
 \frac{\partial}{\partial \phi} \cdot  \frac{\partial}{\partial \phi}
  & \sim & \rho^{\frac{2(1+\kappa)}{1+\kappa+\kappa^2}} \nonumber \\
  \frac{\partial}{\partial z} \cdot  \frac{\partial}{\partial z}
  & \sim & \rho^{\frac{-2 \kappa }{1+\kappa+\kappa^2}}
\end{eqnarray}
and, therefore, for fixed $t$, the length of any segment
corresponding to a finite interval in $z$ with constant $\phi$, and
$\rho$ diverges as we approach $\rho=0$. In particular, one would
have to assign infinite length to any segment of the ``line source''
at $\rho=0$. On the other hand, as $\rho \to \pi/\sqrt{3\Lambda}$,
or, using (\ref{LTeq2c}), as $r \to 0$, we have,
\begin{eqnarray}
\label{Kill02}
 \frac{\partial}{\partial \phi} \cdot  \frac{\partial}{\partial \phi}
  & \sim & r^{\frac{2(2\kappa+1)(\kappa-1)}{3(1+\kappa+\kappa^2)}} \nonumber \\
  \frac{\partial}{\partial z} \cdot  \frac{\partial}{\partial z}
  & \sim & r^{\frac{2 (\kappa+2)(2\kappa+1) }{3(1+\kappa+\kappa^2)}}
\end{eqnarray}
and in this case the length of any segment corresponding to a finite
interval of $\phi$, with constant $\rho$ and $z$, (assuming $\kappa
\geq 0$), diverges as $\rho \to \pi/\sqrt{3\Lambda}$, only if
$\kappa < 1$. Thus, the ``line source'' at $\rho =
\pi/\sqrt{3\Lambda}$, has infinite length only for the range $0 \leq
\kappa \leq 1$.

We can gain some more insights into the physical meaning of the
singularities in the Linet - Tian metric by noticing that,
generally, in the limit $\Lambda =0$ the Linet - Tian solution
reduces to a form of the Levi - Civita metric:
\begin{equation}\label{LCeq1}
ds^2 = -\rho^{\frac{2\kappa(1+\kappa)}{1+\kappa+\kappa^2}}dt^2
+\rho^{\frac{-2\kappa}{1+\kappa+\kappa^2}} dz^2
+\rho^{\frac{2(1+\kappa)}{1+\kappa+\kappa^2}}d\phi^2+ d\rho^2
\end{equation}
But we can also check that (to leading order), (\ref{LCeq1})
corresponds to the limit $\rho \to 0$, for fixed $\Lambda$, so that
the singularity for $\rho \to 0$ of the Linet - Tian metric has the
same nature as that of the Levi - Civita metric. For this latter
metric, in the range $0 \leq \kappa < +\infty$,  $\kappa$ is related
to the mass per unit length of a possible regular material source
that replaces the singularity, and makes the metric regular for
$\rho=0$. For this reason we might consider restricting $\kappa$ to
that range. But we must recall here the equivalent roles played by
$\partial/\partial \phi$, and $\partial/\partial z$, given by the
map $\kappa \to \eta$, and its inverse. In accordance with
(\ref{LTeq2d}), $\eta \geq 0$ only if $\kappa \leq 1$. Since $\eta
\geq 0$ is required to interpret it in terms of a mass per unit
length, in what follows we shall restrict to the range,
\begin{equation}
\label{rangekappa}
    0 \leq \kappa \leq 1
\end{equation}

It is clear from (\ref{LTeq4}) that in all cases at least one of the
$p_i<0$. This, on account of (\ref{LTeq2}), implies that at least
one of the metric coefficients diverges either for $\rho \to 0$, or
$\rho \to \pi/\sqrt{3 \Lambda}$. It is instructive to compute the
Kretschmann scalar $\mathbf{K}$, given by $\mathbf{K} = R^{\alpha
\beta\mu\nu} R_{\alpha \beta\mu\nu}$ corresponding to (\ref{LTeq1}),
to see the effect these divergences have on the structure of the
space time. It is given by,
\begin{eqnarray}\label{Kret1}
 \mathbf{K} & = & \frac{16 \Lambda^2}{3 \sin^4(\sqrt{3
 \Lambda}\rho)} \left[4 \cos^4\left(\frac{\sqrt{3 \Lambda}
 \rho}{2}\right)\left[1 +2 \sin^4\left(\frac{\sqrt{3
 \Lambda}\rho}{2}\right)\right] \right. \nonumber \\
 & & \left. -\frac{(\kappa-1)^2(2
 \kappa+1)^2 (\kappa+2)^2}{(1+\kappa+\kappa^2)^3}
 \cos\left(\sqrt{3\Lambda}\rho\right)\right]
\end{eqnarray}

Near $\rho  =0$ this admits the expansion,
\begin{eqnarray}\label{Kret2}
 \mathbf{K} & = & \frac{16\kappa^2(1+\kappa)^2}{(1+\kappa+\kappa^2)^3} \rho^{-4}
    + \frac{8 \Lambda\kappa^2(1+\kappa)^2}{(1+\kappa+\kappa^2)^3} \rho^{-2} \nonumber \\
 & & +\frac{2 \Lambda^2(36 \kappa^3+43
 \kappa^2+43\kappa^4+30\kappa^5+30\kappa+10 +10 \kappa^6}{5 (1
 +\kappa+\kappa^2)^3}+ {\cal{O}}\left(\rho^2\right),
\end{eqnarray}
and, therefore, the leading divergence is independent of $\Lambda$
and coincides, as expected from the above discussion on the limit of
the Linet - Tian metric as $\rho \to 0$, with the corresponding
singularity for the Levi - Civita metric.

On the other hand, in the limit $\rho = \pi/\sqrt{3\Lambda}$,
setting again $\rho = \pi/\sqrt{3\Lambda}-r$, near $r=0$ we have,
\begin{eqnarray}\label{Kret3}
 \mathbf{K} & = & \frac{16 (\kappa-1)^2(2\kappa+1)^2(\kappa+2)^2}{27 (1+\kappa+\kappa^2)^3} r^{-4}
    + \frac{8 \Lambda(\kappa-1)^2(2\kappa+1)^2(\kappa+2)^2}{27(1+\kappa+\kappa^2)^3} r^{-2} \nonumber \\
 & & +\frac{2 \Lambda^2(202 \kappa^6+606
 \kappa^5+1671\kappa^4+2332\kappa^3+1671\kappa^2+606 \kappa+202}{135 (1
 +\kappa+\kappa^2)^3}+ {\cal{O}}\left(r^2\right).
\end{eqnarray}
or, in terms of $\eta$,
\begin{eqnarray}\label{Kret5}
 \mathbf{K} & = & \frac{16\eta^2(1+\eta)^2}{(1+\eta+\eta^2)^3} r^{-4}
    + \frac{8 \Lambda\eta^2(1+\eta)^2}{(1+\eta+\eta^2)^3} r^{-2} \nonumber \\
 & & +\frac{2 \Lambda^2(36 \eta^3+43
 \eta^2+43\eta^4+30\eta^5+30\eta+10 +10 \eta^6}{5 (1
 +\eta+\eta^2)^3}+ {\cal{O}}\left(r^2\right),
\end{eqnarray}
which is identical to (\ref{Kret2}), but with $\rho$ replaced by $r$
and $\kappa$ by $\eta$, and, therefore, we find the same structure
of the singularities (in $\mathbf{K}$), at both $\rho=0$ and
$\rho=\pi/\sqrt{3 \Lambda}$. This is of course in complete agreement
with the properties of the Linet - Tian metric analyzed above.

The fact that the singularity for $\rho = \pi/\sqrt{3 \Lambda}$ of
the Linet - Tian metric with positive $\Lambda$ has this special
character \cite{wang}, was already noticed by Griffiths and
Podolsky. \cite{griffiths}. In their words, they considered this
case of the Linet - Tian metric as ``apparently cylindrical''. This
is because, as already indicated, as we approach the limit $\rho=
\pi/\sqrt{3 \Lambda}$, we find again a Levi - Civita metric, but
this time with the roles of $\partial/\partial \phi$, and
$\partial/\partial z$ interchanged, and the metric corresponds to
the space time of a line source extended in the $\phi$ direction,
with $\partial/\partial z$ the rotational Killing vector around the
line source. The resulting space time has, for $0 <\rho< \pi/\sqrt{3
\Lambda}$, a ``toroidal'' symmetry, where the orbits of
$\partial/\partial \phi$, and $\partial/\partial z$, are orthogonal
and of finite length. This feature of the Linet - Tian metric with
$\Lambda >0$ was used in \cite{griffiths} to construct extensions of
the metric by matching it to an appropriate Einstein space. Here we
shall be interested in the (linear) stability of the Linet - Tian
metric metric under gravitational perturbations, restricting
$\kappa$ to the range $0 \leq \kappa \leq 1$, (and therefore we also
have $1\geq \eta \geq0$), where we have a simpler physical
interpretation for both singularities.

\section{A new form of the metric.}

It will be convenient, for the analysis of the linear perturbations
of the Linet - Tian metric, to change the coordinate $\rho$ to a new
coordinate $y$, such that,
\begin{equation}\label{xderho}
    y = \sin^2\left(\sqrt{3 \Lambda}\,\rho/2\right)
\end{equation}
We then have,
\begin{eqnarray}
\label{PQdex}
  P(\rho) &=& \frac{2 \sqrt{y}}{\sqrt{3\Lambda}\sqrt{1-y}} \nonumber \\
  Q(\rho) &=& \frac{2 \sqrt{y} \sqrt{1-y}}{\sqrt{3\Lambda} }
\end{eqnarray}
and (after some constant rescaling of $t$, $z$, and $\phi$), the
metric takes the form,
\begin{eqnarray}\label{LTdex}
    ds^2 & = & -y^{1/3 +p_1/2}(1-y)^{1/3-p_1/2} dt^2 + \frac{1}{3
    \Lambda y(1-y)} dy^2 \nonumber \\
    & & +y^{1/3 +p_2/2}(1-y)^{1/3-p_2/2} dz^2+y^{1/3 +p_3/2}(1-y)^{1/3-p_3/2} d\phi^2
\end{eqnarray}
where $y$ is restricted to the range $0\leq y \leq 1$. This is the
form that will be used as the unperturbed metric in the rest of the
paper.

\section{Gauge ambiguities.}

We may write the general linear perturbation of the Linet - Tian
metric in the form,
\begin{equation}\label{gau01}
     g_{\mu\nu}(t,y,z,\phi) =  g_{\mu\nu}^{(0)}(y)+\epsilon  h_{\mu\nu}(t,y,z,\phi)
\end{equation}
where $g_{\mu\nu}^{(0)}(y)$ is the (unperturbed) Linet-Tian metric
(\ref{LTdex}), and $\epsilon$ is an auxiliary parameter, that will
be used to keep track of the linearity of the perturbations.
$h_{\mu\nu}$ represent the most general perturbation. We notice,
however, that since $\partial_t$, $\partial_z$, and
$\partial_{\phi}$ are Killing vectors of $g_{\mu\nu}^{(0)}$, we may
restrict to perturbations of the form,
\begin{equation}\label{gau03}
    h_{\mu\nu}(t,y,z,\phi) = e^{i(\Omega t - k z -\ell \phi)}
    f_{\mu\nu} (y)
\end{equation}

In this paper, however, and because of the indicated relations
between $\partial_z$, and $\partial_{\phi}$, an also for simplicity,
we will consider only the case $\ell=0$, so that the perturbations
will depend only on $(t,x,z)$.  The resulting general perturbed
metric, however, is still subject to gauge ambiguities, resulting
from the fact that we can change coordinates in such a way that the
form of the metric is maintained. More explicitly, consider new
coordinates $(T,Y,Z,\Phi)$, such that,
\begin{eqnarray}
\label{gau05}
  t &=& T + \epsilon \; e^{i(\Omega T - k Z )} Q_T(Y) \nonumber \\
   y &=& Y + \epsilon \; e^{i(\Omega T - k Z)} Q_X(Y) \nonumber \\
   z &=& Z + \epsilon\; e^{i(\Omega T - k Z )} Q_Z(Y) \nonumber \\
   \phi &=& \Phi + \epsilon\; e^{i(\Omega T - k Z)} Q_{\Phi}(Y)
\end{eqnarray}
where the $Q_A$ are arbitrary functions. Then if we write the
general perturbed metric in the form,
\begin{eqnarray}
\label{gau07}
  ds^2 &=& -\frac{y^{1/3+p_1/2}}{(1-y)^{p_1/2-1/3}}\left(1+\epsilon e^{i(\Omega t - k z)}F_1(1)\right) dt^2
        +\frac{1}{3  \Lambda y(1-y)}\left(1+\epsilon e^{i(\Omega t - k z)}F_2(y)\right) dy^2  \nonumber \\
   & & +\frac{y^{1/3+p_2/2}}{(1-y)^{p_2/2-1/3}}\left(1+\epsilon e^{i(\Omega t - k z)}F_3(y)\right) dz^2
   +\frac{y^{1/3+p_3/2}}{(1-y)^{p_3/2-1/3}}\left(1+\epsilon e^{i(\Omega t - k z)}F_4(y)\right) d\phi^2 \nonumber \\
   && +2\epsilon e^{i(\Omega t - k z)}F_5(y)dt dy+2\epsilon e^{i(\Omega t - k z)}F_6(y)dt dz
   +2\epsilon e^{i(\Omega t - k z)}F_7(y)dz dy \nonumber \\
  && +2\epsilon e^{i(\Omega t - k z)}F_8(y)dt d\phi+2\epsilon e^{i(\Omega t - k
  z)}F_9(y)dy
  d\phi
   +2\epsilon e^{i(\Omega t - k z)}F_{10}(y)dz d\phi\;,
\end{eqnarray}
we find that under the transformation (\ref{gau05}), again to linear
order in $\epsilon$, we get a new metric of the same form as
(\ref{gau07}), but with new functions $\widetilde{F}_i$, related to
the old $F_i$ by,
\begin{eqnarray}
\label{gau09}
  \widetilde{F}_1(Y) &=& F_1(Y)  +2 i \Omega Q_T(Y)-\frac{(4 Y-2-3p_1)Q_Y(Y)}{6 Y (1-Y)} \nonumber \\
  \widetilde{F}_2(Y) &=& F_2(Y)+\frac{(2 Y-1) Q_Y(Y)}{Y(1-Y)}+2 \frac{d Q_Y}{dY} \nonumber \\
  \widetilde{F}_3(Y) &=& F_3(Y) +\frac{(3 p_2-4Y+2) Q_Y(Y)}{6 Y (1-Y)}-2 i k Q_Z(Y)\nonumber \\
 \widetilde{F}_4(Y) &=& F_4(Y) +\frac{(2 + 3 p_3 -4 Y) Q_Y(Y)}{6 Y (1-Y)} \nonumber \\
 \widetilde{F}_5(Y) &=& F_5(Y) +\frac{ i \Omega Q_Y(Y)}{3\Lambda Y(1-Y)} -Y^{1/3+p_1/2}
(1-Y)^{1/3-p_1/2} \frac{dQ_T}{dY}\nonumber \\
 \widetilde{F}_6(Y) &=& F_6(Y) +i \Omega Y^{1/3+p_2/2}(1-Y)^{1/3-p_2/2} Q_Z(Y) \nonumber \\
  & &+i k Y^{1/3+p_1/2}(1-Y)^{1/3-p_1/2} Q_T(Y)  \nonumber \\
 \widetilde{F}_7(Y) &=& F_7(Y)
  - \frac{ i k Q_Y(Y)}{3 \Lambda Y(1-Y)} +
 Y^{p_2/2+1/3} (1-Y)^{1/3-p_2/2} \frac{dQ_Z}{dY} \nonumber \\
 \widetilde{F}_8(Y) &=& F_8(Y) +i \Omega  Y^{1/3+p_3/2} (1-Y)^{1/3-p_3/2} Q_{\Phi}(Y) \nonumber \\
 \widetilde{F}_9(Y) &=& F_9(Y) +Y^{1/3+p_3/2} (1-Y)^{1/3-p_3/2} \frac{dQ_{\Phi}}{dY}\nonumber \\
 \widetilde{F}_{10}(Y) &=& F_{10}(Y) -i k  Y^{1/3+p_3/2} (1-Y)^{1/3-p_3/2} Q_{\Phi}(Y)
\end{eqnarray}

Notice that (\ref{gau09}) actually separates into two groups. The
first contains $\widetilde{F}_i,\;(i = 1..7)$, and $Q_T$, $Q_Y$, and
$Q_Z$, while the other contains $\widetilde{F}_i,\;(i = 8,9,10)$,
and only $Q_{\Phi}$. The two groups transform independently. This is
in correspondence with the fact that the linearized Einstein
equations for the perturbed metric split into two sets, one that
couples the $F_i$, with $i = 1..7$ with each other, and a separate
one that couples only $F_8$, $F_9$, and $F_{10}$ with each other.
For reasons to be discussed below, we shall call the latter the
``non diagonal'' case, and the former the ``diagonal case''. In this
paper we will concentrate in the ``diagonal case''. The ``non
diagonal case'', together with several other interesting properties
of the linear perturbations of the Linet - Tian metric will be
considered in a separate paper.

\section{The diagonal case.}

As we have already indicated, the Einstein equations couple only
$F_1$ ,$F_2$, $F_3$, $F_4$, $F_5$, $F_6$, and $F_7$ to each other,
but leave as a separate set $F_8$, $F_9$, and $F_{10}$. Going back
to (\ref{gau09}), it is clear that we can always choose $Q_Y$,
$Q_T$, and $Q_Z$ such that $\widetilde{F}_5=0$, $\widetilde{F}_6=0$,
and $\widetilde{F}_7=0$. (Notice that at this stage we may replace
$Y \to y$, $T \to t$, and $Z\to z$ without any ambiguity). This
implies that without loss of generality we may restrict to the
``diagonal'' case where only $F_1$, $F_2$, $F_3$, and $F_4$ are non
vanishing. This choice is consistent with the equations of motion
but it is not free from gauge ambiguities. This is because, in
accordance with (\ref{gau09}), a coordinate transformation with the
$Q_i$ of the form,
\begin{eqnarray}
\label{diag01}
  Q_t(y) &=&  y^{p_2/4-p_1/4}(1-y)^{p_1/4-p_2/4} \Omega Q_0 \nonumber \\
  Q_y (y) &=& \frac{3 i}{4} y^{1/3-p_3/2} (1-y)^{1/3+p3/4} \Lambda Q_0 \\
  Q_z(y)  &=&   y^{p_1/4-p_2/4}(1-y)^{p_2/4-p_1/4} k Q_0 \nonumber
\end{eqnarray}
where, $Q_0$ is an arbitrary constant, leaves the diagonal form
invariant. This, as shown below, has some important consequences
that will be relevant in the analysis of the resulting equations of
motion.\\

We must remark at this point that there are other choices of gauge,
that is, of the non vanishing $F_i$, that are essentially free of
gauge ambiguities. The problem with those choices is that they lead
to equations that are considerably more complicated and difficult to
handle than the ``diagonal'' choice made for our analysis, and for
this reason they were not considered here.\\

Consider now the linearized Einstein equations. These can be written
in the form,
\begin{equation}\label{diag03}
    F_2(y)=-F_4(y),
\end{equation}
\begin{equation}
\label{diag04a}
  \frac{dF_1}{dy} +\frac{dF_4}{dy} +\frac{p_1-p_2}{4 y (1-y)} F_1
   -\frac{8 y+9 p_2 -4 +3 p_1}{12 y (1-y)} F_4  = 0,
\end{equation}

\begin{equation}
\label{diag04b}
  \frac{dF_3}{dy} +\frac{dF_4}{dy} -\frac{p_1-p_2}{4 y (1-y)} F_3
   -\frac{8 y+9 p_1 -4 +3 p_2}{12 y (1-y)} F_4  = 0,
\end{equation}
and,
\begin{eqnarray}
\label{diag06} {\frac {dF_4}{dy}} & = & {\frac {{\Omega}^{2} \left(
1-y \right) ^{p_1/2-1/3} \left( F_{ {3}} +F_{{4}}  \right)
}{2\Lambda\,{y}^{1/3+ p_1/2}
 \left( 2-4 y+3 p_3 \right) }}
 -{\frac {{k}^{2} \left( 1-y
 \right) ^{p_2/2-1/3}  \left( F_{{1}}
  +F_{{4}}  \right) }{2\Lambda {y}^{p_2/2+1/3}
 \left( 2-4 y+3 p_3 \right) }} \nonumber \\ &&
 +{\frac { \left( {p_1}-{
p_2} \right)  \left( - \left( 8\,y-4+3 p_1 \right) F_{{1}}
  +F_{{3}}   \left( 8y-4+3 p_2
 \right)  \right) }{8 \left( -1+y \right) y \left( 2-4 y+3 p_3
 \right) }} \nonumber \\ &&
 +{\frac { \left( 32 {y}^{2}- \left( 32+120 p_3
 \right) y+60 p_3+45 p_1 p_2+44 \right) F_{{4}}
  }{24 \left( 4y-2-3 p_3 \right) y \left( 1-y
 \right) }}
\end{eqnarray}

Clearly, this system can also be written in the form $dF_i/dy =
f_i(y,F_1,F_3,F_4)$, where the functions $f_i$ are linear in the
$F_i$. Since $4y-3p_3-2 = 4y-6(\kappa+1)/(1+\kappa+\kappa^2) < 0$ in
$0 < y < 1$, the $y$-dependent coefficients of the $F_i$  are
regular in $0 < y <1$, but singular, in general, for both $y=0$ and
$y=1$. This result implies that the general solution of the system
(\ref{diag04a},\ref{diag04b},\ref{diag06}) can be written as a
linear combination of three appropriately chosen linearly
independent solutions, which are {\em regular}, i.e., non singular,
in $0 < y <1$, but may be singular at either or both $y=0$, and
$y=1$. One of these solutions can be obtained immediately replacing
the $F_i$ by their purely gauge dependent part, given by
(\ref{gau09}), with the $Q_i(y)$ given by (\ref{diag01}), and
setting the $F_i=0$ on the right hand side of (\ref{diag01}).
Namely, the set,
\begin{eqnarray}
\label{diag08}
  F_1(y) &=& \frac{(3p_1+2-4y)y^{p_1/4+p_2/4-1/3}(p_1-p_2)\Lambda Q_0}
          {(1-y)^{p_1/4+p_2/4+2/3}} +\frac{16 y^{p_1/4-p_2/4} \Omega^2
          Q_0}{(1-y)^{p_1/4-p_2/4}} \nonumber \\
  F_3(y) &=&  \frac{(3p_2+2-4y)y^{p_1/4+p_2/4-1/3}(p_1-p_2)\Lambda Q_0}
          {(1-y)^{p_1/4+p_2/4+2/3}} +\frac{16 y^{p_2/4-p_1/4} k^2
          Q_0}{(1-y)^{p_2/4-p_1/4}} \nonumber \\
    F_4(y) &=& \frac{(2-4y+3p_3)(p_1-p_2)\Lambda}
    { y^{2/3+p_3/4}(1-y)^{2/3-p_3/4}} Q_0
\end{eqnarray}
where $Q_0$ is a constant, is a pure gauge solution of the system
(\ref{diag04a},\ref{diag04b},\ref{diag06}), that can always be
removed by an appropriate coordinate transformation. Notice that
this solution is regular for $0 < y < 1$, as indicated, but it is
divergent both for $y \to 0$ and $y \to 1$.\\

If we consider the system in more detail, we find that, besides
(\ref{diag08}), we have two other independent solutions, one of
which, near $y=0$, behaves as,
\begin{eqnarray}
\label{diag10}
  F_1(y) & \simeq &  -\frac{2+4\kappa+\kappa^2}{\kappa(2+\kappa)}
   c_0 + a_1 y^{\frac{1}{1+\kappa+\kappa^2}} \nonumber \\
  F_3(y) & \simeq &  -\frac{\kappa^2-2}{\kappa(2+\kappa)} c_0
  + b_1 y^{\frac{1}{1+\kappa+\kappa^2}} \nonumber \\
    F_4(y) & \simeq &   c_0 + c_1 y^{\frac{1}{1+\kappa+\kappa^2}}
\end{eqnarray}
plus higher order terms, where $c_0$ is an arbitrary constant, and
\begin{eqnarray}
\label{diag12}
  a_1 & = &  \frac{(2+\kappa)(\kappa^2-2+2\kappa)(1+\kappa+\kappa^2)^2
   \Omega^2 c_0}{3\kappa(\kappa^2+2\kappa+4)\Lambda} \nonumber \\
   b_1 & = &  \frac{(\kappa-2)(\kappa^2+2+2\kappa)(1+\kappa+\kappa^2)^2
   \Omega^2 c_0}{3\kappa(\kappa^2+2\kappa+4)\Lambda} \nonumber \\
     c_1 & = &  \frac{(2-2\kappa-\kappa^2)(\kappa^2+2+2\kappa)
     (1+\kappa+\kappa^2)^2\Omega^2 c_0}{3(2+\kappa)(\kappa^2+2\kappa+4)\Lambda}
\end{eqnarray}
and, therefore, the $F_i$ approach a finite limit as $y \to 0$, but
with divergent derivatives in that limit, because
$(1+\kappa+\kappa^2)^{-1} < 1$, for $\kappa >0$.

For the other solution, near $y=0$, we have,
\begin{eqnarray}
\label{diag14}
  F_1(y) & \simeq &  -\frac{2+4\kappa+\kappa^2}{\kappa(2+\kappa)}c_2
   \ln(y)+\frac{66\kappa+80 \kappa^2+47\kappa^3+13\kappa^4+20+2\kappa^5}
   {\kappa^3(2+\kappa)^2} c_2 \nonumber \\
  F_3(y) & \simeq &  \frac{2-\kappa^2}{\kappa(2+\kappa)}c_2
  \ln(y)+\frac{34\kappa+16\kappa^2-3\kappa^3-5\kappa^4+20-
  2\kappa^5}{\kappa^3(2+\kappa)^2} c_2  \nonumber \\
    F_4(y) & \simeq &   c_2
    \ln(y)+\frac{2\kappa^2+5\kappa+5}{\kappa^2} c_2
\end{eqnarray}
where $c_2$ is an arbitrary constant, plus terms that vanish as $y
\to 0$, and, therefore, the $F_i$ diverge as $\ln(y)$.

Similarly, near $y=1$, we have a solution that behaves as,
\begin{eqnarray}
\label{diag16}
  F_1(y) & \simeq &  \frac{\kappa^2-8 \kappa-2}{3 \kappa (2+\kappa)} c_3
   +a_4 \frac{k^2}{\Lambda} c_3(1-y)^{\frac{(1-\kappa)^2}{3(1+\kappa+\kappa^2)}}
    \nonumber \\
  F_3(y) & \simeq &  \frac{2-7\kappa^2-4\kappa}{3 \kappa (2+\kappa)} c_3
  +b_4 \frac{k^2}{\Lambda} c_3(1-y)^{\frac{(1-\kappa)^2}{3(1+\kappa+\kappa^2)}}
   \nonumber \\
    F_4(y) & \simeq &    c_3
    +c_4 \frac{k^2}{\Lambda} c_3(1-y)^{\frac{(1-\kappa)^2}{3(1+\kappa+\kappa^2)}}
\end{eqnarray}
plus higher order terms, $c_3$ is an arbitrary constant, and $a_4$,
$b_4$, and $c_4$ are constants that depend only on $\kappa$. For the
other independent solution $F_1$, $F_3$ and $F_4$ diverge as
$\ln(1-y)$ as $y \to 1$, but we shall not display their leading
behaviour for simplicity. Thus, we see that the system has solutions
that are well behaved, i.e., do not diverge, at either $y=0$ or
$y=1$.\\

What this means is that if we consider a solution that behaves as
(\ref{diag10}) near $y=0$, then, in general, as we approach $y=1$,
it will behave as a linear combination of the three linearly
independent solutions characterized by their behaviour near $y=1$,
and, therefore, it will diverge for $y \to 1$. As discussed, for
instance in \cite{glei1} or \cite{glei2}, we should, in principle,
consider only as appropriate those solutions of the perturbation
equations such that the $F_i$ do not diverge either at $y=0$ or
$y=1$. Since solutions of the system can only be obtained
numerically, one might then try to impose this condition at say
$y=0$, and, for fixed $\kappa$ and $k$, look for possible values of
$\Omega$, such that the solution is also finite as we approach
$y=1$. Unfortunately, because of the gauge ambiguities contained in
the system, this simple ``shooting'' procedure fails to provide the
required solutions. What is required here is a {\em gauge invariant
function} that carries the physical properties of the perturbations,
and satisfies the finiteness requirements, while the $F_i$
themselves may still contain gauge dependent divergent components.
This problem is considered in the next Section.

\section{Gauge invariant formulation.}

Gauge invariant functions may be constructed in general as a linear
combinations of the $F_i(y)$. Let us call $F^{g}_i(y)$ the solutions
given by (\ref{diag08}), then, a suitable example, is the function,
\begin{equation}\label{diag18}
    W(y) =
    {\cal{K}}(y)\left[F^{g}_3(y)F_4(y)-F^{g}_4(y)F_3(y)\right]
\end{equation}
where ${\cal{K}}$ is an arbitrary function of $y$. If we choose,
\begin{equation}\label{diag20}
    {\cal{K}}(y) = C_K \frac{y^{2/3+p3/4}}{(1-y)^{p3/4-2/3}},
\end{equation}
after adjusting the constant $C_K$, we get,
\begin{eqnarray}
\label{diag22}
  W_1 \left( y \right) & = &-\Lambda \left( 2-4y+3 p_3 \right)  \left(p_1-p_2 \right) F_{{3}}
  \nonumber  \\ &  &+ \left( \Lambda \left( 2-4y+3p_2 \right)  \left(
p_1-p_2 \right) +16 \left( 1-y \right) ^{2/3+p_2/2}
{y}^{p_2/2+2/3}{k}^{2} \right) F_{{4}}
\end{eqnarray}
Notice that, since,
\begin{equation}\label{diag24}
     y^{2/3-p_2/2} (1-y)^{2/3+p_2/2} =
     y^{\frac{(1+\kappa)^2}{1+\kappa+\kappa^2}}
     (1-y)^{\frac{(1-\kappa)^2}{3 (1+\kappa+\kappa^2)}},
\end{equation}
the coefficients of $F_3$, and $F_4$ are finite both for $y \to 0$
and $y \to 1$. In particular, near $y=0$, for the solution
(\ref{diag10}) we have,
\begin{equation}\label{diag26}
 W_1(y) \simeq
 \frac{72(1+\kappa)(2+2\kappa+\kappa^2)}{(1+\kappa+\kappa^2)^3}
 \Lambda c_0
 -\frac{24(1+\kappa)(2+2\kappa+\kappa^2)}{(1+\kappa+\kappa^2)}
 \Omega^2 c_0 y^{\frac{1}{1+\kappa+\kappa^2}},
\end{equation}
and, near $y=1$, for the solution (\ref{diag16}),
\begin{equation}\label{diag28}
 W_1(y) \simeq \frac{8(1-\kappa)(5 \kappa^2+2\kappa+2)(2\kappa+1)^3}{3
 (1+\kappa+\kappa^2)^3} \Lambda c_3 + d_4 c_3 k^2
 (1-y)^{\frac{(1-\kappa)^2}{3(1+\kappa+\kappa^2)}}
\end{equation}
plus higher order terms, and where $d_4$ is a constant that depends
only on $\kappa$. Thus, $W_1$ is well defined and finite for data
that satisfies the finite boundary conditions (\ref{diag10}),
(\ref{diag16}).\\

But a crucial property of $W_1$ is that it is not only gauge
invariant, but it is also a {\em master variable}, in the sense that
the full perturbation can be reconstructed from $W_1$. This can be
seen as follows. First, we solve(\ref{diag22}) for $F_3$ in terms of
$W_1$, and, $F_4$,
\begin{equation}
\label{diag30}
  F_3(x) = \frac{ {\cal{K}}  F^{g}_3 F_4 -W_1}{ {\cal{K}}F^{g}_4}
\end{equation}

Replacing (\ref{diag30}) in (\ref{diag04b}), using the fact that the
$ F^{g}_i$ are solutions of (\ref{diag04b}), and rearranging terms
we find,
\begin{eqnarray}
\label{diag31}
  {\frac {d}{dy}} \left( \frac {F_4  }{F^{g}_4}
  \right)& = &{\frac {{W_1}
 \left( p_2-p_1 \right) }{4 F^{g}_4 {\cal{K}}
   \left( F^{g}_3 + F^{g}_4
   \right) y \left( 1-y \right) }}+
  \frac{1}{F^{g}_3 + F^{g}_4}
   {\frac {d}{dy}}
 \left( {\frac {{W_1} }{F^{g}_4 {\cal{K}}  }} \right),
\end{eqnarray}
which implies,
\begin{eqnarray}
\label{diag32}
  F_4(y)& = & F^{g}_4 \int_0^y{\left[{\frac {
 \left( p_2-p_1 \right) {W_1}}{4 F^{g}_4 {\cal{K}}
   \left( F^{g}_3 + F^{g}_4
   \right) y \left( 1-y \right) }}+
  \frac{1}{(F^{g}_3 + F^{g}_4)}
   {\frac {d}{dy}}
 \left( {\frac {{W_1} }{F^{g}_4 {\cal{K}}  }} \right)\right] dy} \nonumber \\
 && + C F^{g}_4(y),
\end{eqnarray}
where $C$ is an arbitrary constant, and, therefore, we can express
$F_4$ entirely in terms of $W_1$, and the already known pure gauge
solutions.

Using the expressions for $F_3$, and $F_4$ we may also obtain an
expression for $F_1(y)$, in terms of $W_1(y)$, but it turned out to
be more useful for the derivations to solve (\ref{diag06}) for
$F_1(y)$. This is given by,

\begin{eqnarray}
\label{diag32a}
 F_{{1}}   & = &{\frac {16\mu \Lambda y \left( 1-y
 \right)  \left( 2 \mu y-3-3 \kappa \right)}  {{ A_1}}}\frac {dF_4}{dy}
 +{\frac {4{A_4}\,F_{{4}} }{3{A_1}}}  \nonumber \\ & &
 +4 \frac{  4\, \left( 1-y \right) ^{{
\frac { \left( 2\,\kappa+1 \right)
^{2}}{3\mu}}}{y}^{{\mu}^{-1}}{\mu}^{ 2}{\Omega}^{2}-\kappa\, \left(
\kappa+2 \right)  \left( 4\,\mu\,y-3\,
 \left( 1+\kappa \right) ^{2} \right) \Lambda } {{A_1}} F_3
\end{eqnarray}
where $\mu =1+\kappa+\kappa^2$,
\begin{equation}\label{diagA1}
 A_1(y)=16\,{\mu}^{2} \left( 1-y \right) ^{{\frac { \left(
\kappa-1 \right) ^{2}}{3\mu}}}{y}^{{\frac { \left( 1+\kappa \right)
^{2 }}{\mu}}}{k}^{2}-4\,\kappa\, \left( \kappa+2 \right)  \left(
4\,\mu\,y -3 \right) \Lambda
\end{equation}
and,
\begin{eqnarray}
\label{diagA4}
 A_4(y)& = &12 \left( 1-y \right) ^{{\frac { \left(
2\,\kappa+1
 \right) ^{2}}{3\mu}}}{y}^{\frac{1}{\mu}}{\mu}^{2}{\Omega}^{2}-12\,{\mu}^{
2} \left( 1-y \right) ^{{\frac { \left( \kappa-1 \right) ^{2}}{3
\mu}}}{y}^{{\frac { \left( 1+\kappa \right) ^{2}}{\mu}}}{k}^{2}
\nonumber \\ && -
 \left( 8{y}^{2}{\mu}^{2}+12\mu \left( {\kappa}^{2}-4\kappa-4
 \right) y+72\kappa-9{\kappa}^{4}+36-18{\kappa}^{3}+18{\kappa}
^{2} \right) \Lambda
\end{eqnarray}\\

Thus, as indicated, we have succeeded in expressing the full
diagonal perturbation in terms of the master function $W_1$. The
resulting expressions, nevertheless, still contain the gauge
ambiguities. In fact, going back to (\ref{diag32}), we can see as
expected, that $F_4$ reduces to $F^{g}_4$ when $W_1(y)=0$, the pure
gauge situation. But, suppose now that we insert in (\ref{diag32})
an appropriate non trivial $W_1(y)$, satisfying the boundary
conditions (\ref{diag26},\ref{diag28}). It is easy to check that if
we also set $C=0$, the resulting $F_4(y)$ satisfies (\ref{diag10})
near $y=0$. But, we can also check that near $y=1$, since the
integral is finite, $F_4(y)$ approaches in general $F^{g}_4(y)$.
There is no contradiction here, it simply means that we cannot
choose a simple gauge where $F_4$ is free of $F^{g}_4(y)$
``contamination''. This suggests that we look directly for the
equation that $W_1(y)$ should satisfy, when the $F_i$ satisfy their
corresponding equations. This can be achieved going back
(\ref{diag31}), and taking a new $y$-derivative. Solving for $d^2
W_1/dy^2$, and after several replacements, using the evolution
equations for the $F_i$, we finally get the following equation for
$W_1(y)$,
\begin{eqnarray}
\label{diag40}
  -{\frac {d^{2}W_1}{d{y}^{2}}} +{\frac {{4 A_2} }{3 y \left(y -
1 \right) { A_1}}}{\frac {dW_1}{dy}} -{\frac { 4{ A_3}}{ 3y \left(y
-1
 \right) ^{2}\Lambda\,{A_1}}} W_1
  =\frac{ {\Omega}^{2}}{\Lambda
  {y}^{{\frac {2\mu-1}{\mu}}}
   \left( 1-y \right) ^{{\frac {2\,\mu+3}{3\mu}}} }
 W_1
\end{eqnarray}
where $\mu=1+\kappa+\kappa^2$,
\begin{eqnarray}\label{diagA2}
  A_2(y)& = &4 \left( 1-y \right) ^{{\frac { \left( \kappa-1
 \right) ^{2}}{3\mu}}}{y}^{{\frac { \left( 1+\kappa \right) ^{2}}{\mu}}
}\mu \left( 2  \mu y-3 {\kappa}^{2}-9 \kappa-3 \right) {k}^{2}
\nonumber \\ && +3 \kappa \left( \kappa+2 \right)  \left( 3+ \left(
-2+4 \kappa+4{ \kappa}^{2} \right) y \right) \Lambda
\end{eqnarray}
and,
\begin{eqnarray}\label{diagA3}
  A_3(y) & = &-4\,{y}^{{\frac {1+3\,\kappa+{\kappa}^{2}}{\mu}}} \left( 1-y
 \right) ^{{\frac {2 \left( \kappa-1 \right) ^{2}}{3\mu}}}{\mu}^{2}
{k}^{4} \nonumber \\ & & +{y}^{{\frac {\kappa}{\mu}}} \left( 1-y
\right) ^{{\frac {
 \left( \kappa-1 \right) ^{2}}{3\mu}}} \left[ 8{y}^{2}{\mu}^{2}-4\,
\mu \left( 2+8\kappa+5\,{\kappa}^{2} \right) y  \right. \nonumber\\
& & \left.+3\kappa \left( 2\kappa+3 \right)  \left( 2\kappa+1
\right) \left( \kappa+2
 \right)  \right] \Lambda{k}^{2}
 +6\kappa \left( \kappa+2
 \right)  \left( 2\mu -3\right)  \left(y -1
 \right) {\Lambda}^{2}.
\end{eqnarray}\\

We notice now that (\ref{diag40}) has the general form,
\begin{eqnarray}
\label{diag42}
 - \frac{d^2 W_1}{dy^2}+Q_1(y) \frac{d
  W_1}{dy} +Q_2(y) W_1
    =\frac{4}
   {3 \Lambda (1-y)^{\frac{2\mu+3}{3\mu}}
    y^{\frac{2\mu-1}{\mu}}} \Omega^2 W_1
\end{eqnarray}

This may be put in a Schrodinger - like form introducing a new
coordinate $r=r(y)$, and two new functions, $K(y)$, and
$\widetilde{W_1}(r)$, such that,
\begin{equation}\label{diag44}
    W_1(y)=K(y)\widetilde{W_1}\left(r(y)\right)
\end{equation}

Replacing in (\ref{diag42}) we get,
\begin{eqnarray}\label{diag46}
& &-{\frac {d^{2} \widetilde{W_1}}{d{r}^{2}}} -{\frac { \left( 2
\left( {\dfrac {dK}{dy}} \right) {\dfrac {dr}{dy}}
  +K  {\dfrac {d^{2}r}{d{y}^{2}}} -{ Q_1} K  {
\dfrac {dr}{dy}}  \right)   }{K   \left( {\dfrac {dr}{dy}}
   \right) ^{2}}}{\frac {d \widetilde{W}}{dr}}-{\frac {  \left( {\dfrac {d^{2}K}{d{y}^{2}}} -{Q_1}
  {\dfrac {d K}{dy}} -{Q_2}  K   \right) }{K   \left( {
\dfrac {d r}{dy}}  \right) ^{2}}}\widetilde{W_1} \nonumber \\
&&= \frac{4 \left( 1-y \right) ^{-{\frac {2\mu+3}{3\mu}}}{y}^{
{\frac {1-2\mu }{\mu}}} {\Omega}^{2} }{3{\Lambda} \left( {\dfrac {d
r} {dy}} \right) ^{2}} \widetilde{W_1}
\end{eqnarray}

If we impose now that $r(y)$ be a solution of,
\begin{equation}\label{diag48}
{\frac {d r}{dy}} =\frac{2}{\sqrt {3} \left( 1-y
 \right) ^{\frac{2\mu+3}{6\mu}}{y}^{\frac {2\mu-1}{2\mu}}},
\end{equation}
and also that $K(y)$ is a solution of,
\begin{equation}\label{diag50}
2\dfrac {dr}{dy}{\frac {d K}{dy}} +
 \left( {\frac {d^{2}r}{d{y}^{2}}} -{ Q_1} {\frac {dr}{dy}}  \right)K = 0,
\end{equation}
replacing in (\ref{diag42}), we find that $\widetilde{W_1}$
satisfies the Schr\"odinger - like equation,
\begin{equation}\label{diag52}
-{\frac {d^{2} \widetilde{W}}{d{r}^{2}}} +
\mathbf{V}\widetilde{W_1}= \frac{\Omega^2}{\Lambda}\widetilde{W_1},
\end{equation}
where the ``potential'' $\mathbf{V}$ is given by,
\begin{equation}\label{diag54}
\mathbf{V}  ={\frac {2  {\dfrac { d r}{dy}} {\dfrac
{d^{3}r}{d{y}^{3}}} -3 \left( {\dfrac {d^{2}r}{d{y}^{2}}}
 \right) ^{2}+ \left( {\dfrac {d r}{dy}}  \right) ^{2}
 \left(   Q_1{}^{2}-2{\dfrac {d Q_1}{
dy}} +4{ Q_2}   \right) } {4 \left( {\dfrac {dr}{dy}} \right) ^{4}}}
\end{equation}
and, therefore, it is explicitly given as a function of $y$, through
(\ref{diag48}), even if we do not have explicit solutions for either
(\ref{diag48}) or (\ref{diag50}). Actually, in our case we do have
the general solution of (\ref{diag48}),
\begin{equation}\label{diag56}
r \left( y \right) = \frac{2 \mu {y}^{ \frac{1} {\mu
 }}}{\sqrt {3}}\;
{{}_2F_1({\frac{1}{ 2\mu}} ,{\frac {3+2\mu}{6\mu}};{\frac
{1+2\mu}{2\mu}};{y}^{2})} +C,
\end{equation}
where ${}_2F_1(a,b;c;{x}^{2})$ is a hypergeometric function, with
$C$ an arbitrary constant, that we may set equal to zero. We may use
now (\ref{diag56}) to construct a parametric representation of
$\mathbf{V}(y)$. This would in principle allow us, as in similar
quantum  mechanical problems, to carry out  a qualitative analysis
of the possible spectrum of allowed values of the ``eigenvalues''
$\Omega^2/\Lambda$, and therefore obtain information on the
existence of solutions with $\Omega^2 <0$, signalling unstable
solutions of the evolution equations. Unfortunately, in our case,
that is, Eq. (\ref{diag40}), the functions $Q_1$, and $Q_2$ have
vanishing denominators at some point $0 < y_0 < 1$. This is because
irrespective of the value of $k$, the function $A_1$ is continuous
in $0<y<1$, and we have $A_1(0)= -3\kappa(\kappa+2)<0$ and
$A_1(1)=\kappa(\kappa+2)\mu >0$. This vanishing of the denominators
introduces single poles as functions of $y$ in (\ref{diag40}), but,
as can be checked, it implies that $\mathbf{V}(y)$ has a double pole
at the corresponding value of $y$, and, therefore, (\ref{diag52})
cannot be made self adjoint, and the analysis fails. On this account
we need to go back to system
(\ref{diag04a},\ref{diag04b},\ref{diag06}), and analyze it as it
stands.

\section{Setting up the problem.}

Let us go back to (\ref{gau03}). The idea there is that solving the
equations for the $f_{\mu\nu}$, for {\em fixed} $\Omega$, $k$, and
$\ell$, we should get a {\em complete set}, in the sense that one
should be able to express the evolution of an arbitrary perturbation
in the form,
\begin{equation}
\label{com02}
    h_{\mu\nu}(t,y,z,\phi) = \sum_k{ \sum_{\ell}
    {\sum_{\Omega}{{\cal{C}}_{k,\ell,\Omega}\, e^{i(\Omega t - k z -\ell \phi)}
    f_{\mu\nu} (y,\Omega,k,\ell)}}}
\end{equation}
where the coefficients ${\cal{C}}_{k,\ell,\Omega}$ are determined by
the initial data, and, therefore, the central problem is
constructing appropriate sets of functions $f_{\mu\nu}$.

In the previous sections we found that for the diagonal
perturbations $W_1(y)$ is a not only {\em gauge invariant}, but it
is also a {\em master function}, in terms of which we can express
{\em all} the metric coefficients involved in that class of
perturbations. By imposing that $W_1(y)$ must satisfy appropriate
boundary conditions both at $y=0$ and $y=1$, we transform
(\ref{diag40}) in a boundary value problem that determines the
acceptable solutions $W_1(y)$, and associated values  of $\Omega$.
As we have shown, these solutions are all finite, in spite of the
fact that the coefficients in (\ref{diag40}) are singular. Although
our argument is based on its definition in terms of the $F_i$, it is
easy to check that if the singularity is at $y=y_0$, where $y_0$ is
the solution of,
\begin{equation}
 \label{com02a}
 {k}^{2}=\frac{\kappa\,\Lambda\, \left( \kappa+2 \right)  \left( 4\,{y_0}\,\mu-3 \right)}{   \left( 1-{\it y0} \right)
^{{\frac { \left( \kappa-1 \right) ^{2}}{3\mu}}}
 {{y_0}}^{{\frac { \left( 1+\kappa \right) ^{2}}{\mu}}}
 {\mu}^{2}}
\end{equation}
then, in the neighbourhood of $y=y_0$, the {\em general} solution of
(\ref{diag40}), admits a regular expansion of the form,
\begin{eqnarray}
\label{com02b}
  W_1 &=&a_{{0}}+{\frac {a_{{0}} \left( 2\,{y_0}\,\mu-\kappa-2 \right)  \left( -3\,\mu+3+2\,{ y_0}\,\mu
 \right)  \left( y-{ y_0} \right) }{ 2\left( -3-3\,\kappa+2\,{ y_0}
\,\mu \right)  \left( { y_0}-1 \right) {y_0}\,\mu}}\nonumber
\\ && +  a_{{0}} \left( \frac{
 \left( 1-{ y_0} \right) ^{-{\frac { 2\mu+3
}{3\mu}}}{\Omega}^{2}}{{\Lambda}{{
y_0}}^{\frac{2\mu-1}{\mu}}}-{\frac {\kappa \left( \kappa+2 \right)
 \left( 2\mu(y_0-3)+9+3\kappa \right)  \left( 4{y_0
}\mu-3 \right) }{24{{ y_0}}^{2} \left( -3-3\kappa+2{y_0}\mu
 \right)  \left( { y_0}-1 \right) ^{2}{\mu}^{2}}} \right)
 \left( y-{  y_0} \right) ^{2} \nonumber \\ && +a_{{3}}
 \left( y-{ y_0} \right) ^{3}+a_{{4}} \left( y- y_0 \right)
 ^{4}+\dots
\end{eqnarray}
where $a_0$, and $a_3$ are arbitrary constants, $a_4$ is determined
in terms of $a_0$ and $a_3$, and dots indicate higher order terms,
also completely determined in terms of $a_0$, and $a_3$. Notice
that, since $k^2> 0$, we must have $1/4 \leq 3/(4\mu) \leq y_0
<1$.\\

Since the interval $0\leq y \leq 1$ is finite, regularity of
$W_1(y)$ in the interval, plus a ``shooting'' type argument for the
behaviour at $y=1$, starting with, for instance, the boundary
conditions at $y=0$, indicates that the spectrum of allowed values
of $\Omega$ must be discrete. We may then label the solutions with a
discrete index $\lambda$ as, $\left\{ W_1^{(\lambda)},
\Omega_{(\lambda)} \right\}$, but, unfortunately, using only
(\ref{diag40}), it is not at all clear how to obtain other
properties of the set of solutions, such as completeness, or whether
the spectrum is bounded from below. The main problem, as we have
seen, is that the simple attempt to put (\ref{diag40}) in a self -
adjoint form, using $\widetilde{W_1}$, that provided the answer to
those questions in other cases, for instance in \cite{glei2}, fails
here because the ``potential'' has a second order pole. But this is
precisely the situation considered in \cite{dotti}, where it was
shown that one can solve the problem by considering a supersymmetric
pair of (\ref{diag52}). The explicit construction there was in part
made possible by the availability of appropriate exact solutions,
which, in the present situation we do not have. Nevertheless,
because of the formal similarity of both problems, and their
physical nature, it seems reasonable to assume that a construction
similar to that in \cite{dotti} can also be carried out here. This
will be analyzed in a separate study. Here, in the following
sections, we will consider a numerical analysis, based on the system
of equations satisfied by $F_1(y)$, $F_3(y)$, and $F_4(y)$ that
indicates both the existence of unstable modes and a lower bound in
the spectrum of $\Omega^2$.

\section{Numerical analysis.}

There are, in principle, different manners of handling the problem
of a numerical integration of the set
(\ref{diag04a},\ref{diag04b},\ref{diag06}). Since, as indicated, our
main question is, given appropriate boundary conditions at $y=0$ and
$y=1$, are there non trivial, gauge invariant solutions
corresponding to $\Omega^2<0$? We may look for an answer to this
question for given values of $\kappa$ and $k$ by imposing the
regular boundary conditions at either $y=0$ or $y=1$, on the set
(\ref{diag04a},\ref{diag04b},\ref{diag06}), and then analyzing the
solutions that result as we change the values of $\Omega$.

\subsection{$\kappa =1/4$.}

The general case, that is, finding appropriate expressions for a
numerical treatment for general $\kappa$, turns out to be too
complicated, because of the presence of exponents of both $y$, and
$(1-y)$ that are not simple functions of $\kappa$. To make the
discussion more definite, we consider first the case $\kappa=1/4$ in
some detail, and then give several results for the case
$\kappa=1/3$. The special cases $\kappa=0$ and, $\kappa=1$ are
considered in the next Section. For $\kappa=1/4$, the set
(\ref{diag04a},\ref{diag04b},\ref{diag06}), can be written as,
\begin{equation}\label{num02}
 {\frac {d F_1}{dy}}=-{
\frac {dF_4}{dy}}
  -\,{\frac {3 F_{{1}}  }{ 14\left( 1-y \right)
y}}-{\frac { \left( 7-4y \right) F_{{4}}  }{6
 \left( 1-y \right) y}},
\end{equation}
\begin{equation}\label{num04}
 {\frac {d F_3}{dy}}=-{
\frac {dF_4}{dy}}
  -\,{\frac {3 F_{{3}}  }{ 14\left( 1-y \right)
y}}-{\frac { \left( 31-28y \right) F_{{4}}  }{42
 \left( 1-y \right) y}},
\end{equation}
and,
\begin{eqnarray}
\label{num06}
 {\frac {dF_4}{dy}} & = & -\frac{7 \left( F_{{3}}  +F_{{4}}   \right) {\Omega}^{2}}
 {2 {y}^{5/21} \left( 1-y \right) ^{3/7} \left( 7\,y-10 \right){
\Lambda}} + \frac{7{y}^{4/21} \left( F_{{1}}  +F_{{4}} \right)
{k}^{2}}{2 \left( 1-y
 \right) ^{6/7}{\Lambda} \left( 7\,y-10 \right)}
  \nonumber \\&&+{\frac { \left(144 -252 y \right) F_{{1}}  +
 \left( 252 y-225 \right) F_{{3}} + \left( 392{y}^
{2}-2212y+1559 \right) F_{{4}}  }{168 \left( 1-y
 \right) y \left( 7\,y-10 \right) }}
\end{eqnarray}

The first problem in constructing a numerical solution is that we
cannot impose the boundary condition at either $y=0$ or $y=1$
because the coefficients of the equations are singular there. In
this case we may use, if possible, an expansion in appropriate
powers of either $y$ or $(1-y)$ that expresses the required boundary
condition. Consider first the boundary $y=0$. We notice that,
besides integer powers of $y$ we have integer powers of $y^{1/21}$.
We therefore look for an expansion in terms of integer powers of
$y^{1/21}$, which, in this case, takes the form,
\begin{eqnarray}
\label{num08}
  F_1(y) &=& a_0+\frac{5589 \Omega^2 a_0}{18688 \Lambda} y^{16/21}
   -\frac{54 a_0}{581} y + \frac{12177 k^2 a_0}{56875 \Lambda} y^{25/21} + \dots
   \nonumber \\
  F_3(y) &=& -\frac{31}{49}a_0 +\frac{7749 \Omega^2 a_0}{18688 \Lambda} y^{16/21}
   -\frac{594 a_0}{4067} y +\frac{14337 k^2 a_0}{56875 \Lambda}y^{25/21} +\dots
   \nonumber  \\
  F_4(y) &=& -\frac{9}{49}a_0 -\frac{2829 \Omega^2 a_0}{18688
  \Lambda} y^{16/21} -\frac{18 a_0}{4067} y -\frac{7257 k^2
  a_0}{56875 \Lambda} y^{25/21} +\dots
\end{eqnarray}
Actually, for the numerical procedure we carried out the expansion
up to and including the terms in $y^2$, which, as we show below,
provides enough precision. Notice, also, that although the $F_i$
approach finite values, their first derivatives diverge for $y=0$.

Similarly, for $y=1$ we have integer powers of $(1-y)^{1/7}$
\begin{eqnarray}
\label{num10}
  F_1(y) &=& a_{{0}}+{\frac {25}{3}}\,{\frac {{k}^{2}a_{{0}}{(1-y)}^{1/7}}{\Lambda
}}+{\frac {1715}{108}}\,{\frac {{k}^{4}a_{{0}} \left( 1-y \right)
^{2/ 7}}{{\Lambda}^{2}}}+{\frac {84035}{10692}}\,{\frac
{{k}^{6}a_{{0}}
 \left( 1-y \right) ^{3/7}}{{\Lambda}^{3}}} \nonumber \\ &&
 -{\frac {7}{606528}}\,{
\frac {a_{{0}} \left(
588245\,{k}^{8}+10692\,{\Omega}^{2}{\Lambda}^{3}
 \right)  \left( 1-y \right) ^{4/7}}{{\Lambda}^{4}}}
 + \dots
   \nonumber \\
  F_3(y) &=& -\frac{1}{7}a_0 + {\frac {{k}^{2}a_{{0}}({1-y})^{1/7}}{3\Lambda}}-{\frac {245}{108}
}\,{\frac {{k}^{4}a_{{0}} \left( 1-y \right)
^{2/7}}{{\Lambda}^{2}}}-{ \frac {420175}{32076}}\,{\frac
{{k}^{6}a_{{0}} \left( 1-y \right) ^{3/ 7}}{{\Lambda}^{3}}}
\nonumber \\ && -{\frac {245}{6671808}}\,{\frac {a_{{0}} \left(
588245\,{k}^{8}+10692\,{\Omega}^{2}{\Lambda}^{3} \right)  \left( 1-y
 \right) ^{4/7}}{{\Lambda}^{4}}}
 +\dots
   \nonumber  \\
  F_4(y) &=& -\frac{3}{7} a_{{0}}-{\frac {5{k}^{2}a_{{0}}{(1-y)^{1/7}}}{3\Lambda}}+{
\frac {1715}{324}}\,{\frac {{k}^{4}a_{{0}} \left( 1-y \right)
^{2/7}}{ {\Lambda}^{2}}}+{\frac {84035}{3564}}\,{\frac
{{k}^{6}a_{{0}} \left( 1 -y \right) ^{3/7}}{{\Lambda}^{3}}}
\nonumber \\ &&+{\frac {35}{606528}}\,{\frac {a_{{0 }} \left(
588245\,{k}^{8}+10692\,{\Omega}^{2}{\Lambda}^{3} \right)
 \left( 1-y \right) ^{4/7}}{{\Lambda}^{4}}}
 +\dots
\end{eqnarray}
and, again, for the numerical computation we carried out the
expansions up to and including terms of order $(1-y)^{11/7}$. For
the numerical integration we used a Runge - Kutta method, and
(\ref{num08}), or (\ref{num10}) to specify initial values close to
the corresponding boundary. As first check, we compared the
numerical integrations, enforcing the boundary conditions at either
$y=0.0001$ or $y=0.999$, and found a good agreement between the
resulting numerical integrations and the expansions (\ref{num08}),
or (\ref{num10}), sufficiently close to the corresponding boundary
as shown in Figures 1 and 2.

\begin{figure}
\centerline{\includegraphics[height=12cm,angle=-90]{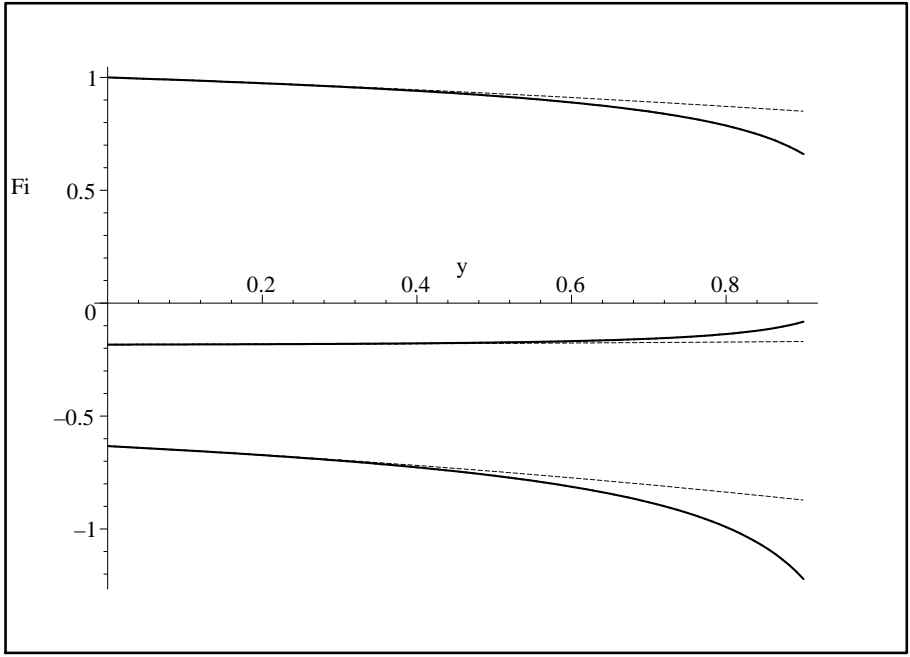}}
\caption{The solid curves correspond to the numerical integration of
the $F_i(y)$, as functions of $y$, enforcing the boundary conditions
(\ref{num08}) at $y=0.0001$. (The upper curve corresponds to $F_1$,
center to $F_4$, and lower to $F_3$). The dotted curves correspond
to the expansions (\ref{num08}). Notice the good agreement for
values of $y$ close to $y=0$, up to $ y \sim 0.2$.}
\end{figure}


\begin{figure}
\vspace{1cm}
\centerline{\includegraphics[height=12cm,angle=-90]{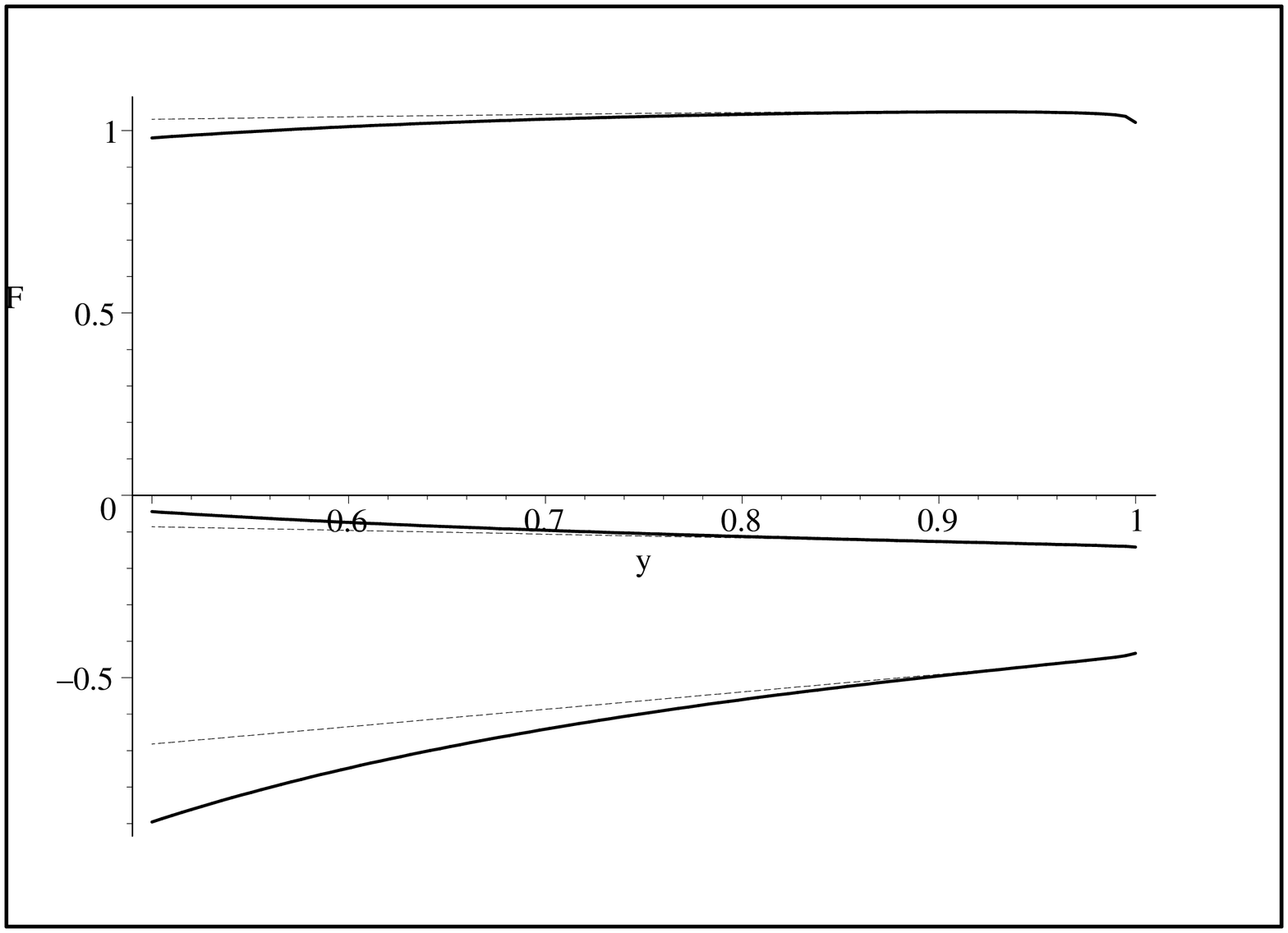}}
\caption{The solid curves correspond to the numerical integration of
the $F_i(y)$, as functions of $y$, enforcing the boundary conditions
(\ref{num10}) at $y=0.999$. (The upper curve to $F_1$, center to
$F_4$, and lower to $F_3$). The dotted curves correspond to the
expansions (\ref{num10}). There is the good agreement for values of
$y$ close $y=1$, up to $ y \sim 0.9$.}
\end{figure}

\vspace{1cm}

The numerically integrated values of the $F_i$ were then used to
compute and plot $W_1(y)$, as a function of $y$. The computations
were carried out separately imposing the regular boundary conditions
at either $y=0$,or $y=1$, keeping fixed $k=0.1$, and $\Lambda=1$,
and changing the value of $\Omega^2$, until a we found a solution
that was regular (by construction) at the end where the regular
initial data was imposed, and such that it would start to diverge in
opposite directions, as we made $\Omega^2$ slightly larger, or
smaller than a certain critical value. This is shown in Figure 3,
where the solid line corresponds $W_1$ for the critical value, which
we identify with the eigenvalue, and we have also indicated with
dotted lines the curves obtained by slightly increasing or
decreasing $\Omega^2$. The solid curve is actually two plots, one
where the regular boundary condition is imposed at $y=0$, and the
other where this is done at $y=1$, both corresponding to $\Omega^2 =
-0.0484...$. To the accuracy of the plot, they are identical,
showing the consistency of the ``shooting method'' used to identify
the required solutions. This was the lowest value we found for
$\Omega^2$. For values lower that this one, the curves
diverge faster and faster as we try lower values for $\Omega^2$.\\

An interesting feature of the numerical integration is the behaviour
of the resulting $F_i$. As indicated, and in accordance with
(\ref{diag32}), even if impose the regular boundary condition at,
say, $y=0$, and find the appropriate value value of $\Omega^2$, so
that $W_1(y)$ is also regular at $y=1$, as in the example of Figure
3, the behaviour of the $F_i$ at $y=1$ will be dominated by the pure
gauge solution. This is illustrated in Figure 4, where the regular
boundary condition was imposed at $y=0$, and the $F_i$ depicted
correspond to the regular solution of Figure 3, but, nevertheless,
the diverging behaviour corresponding to the dominance of the pure
gauge part of the solution is clearly seen near $y=1$.\\

We also looked for critical values larger than the lowest. Figure 5
is plot of $W_1$, for $\kappa=1/2$, $k=0.1$, and $\Lambda=1$,
corresponding to $\Omega^2= 1.030$. As in Figure 3, the plot is a
superposition of the integrations imposing the regular boundary
condition at either $y=0$ or $y=1$, and they coincide within the
accuracy of the plot. Larger values of $\Omega^2$ can be obtained by
the same procedure.

It is important to remark, before closing this subsection, that the
unperturbed metric for $\kappa=1/2$ is isometric to that with
$\kappa=1/4$, and, therefore, an instability for $\kappa=1/4$
implies also an instability for $\kappa=1/2$.

\begin{figure}
\centerline{\includegraphics[height=12cm,angle=-90]{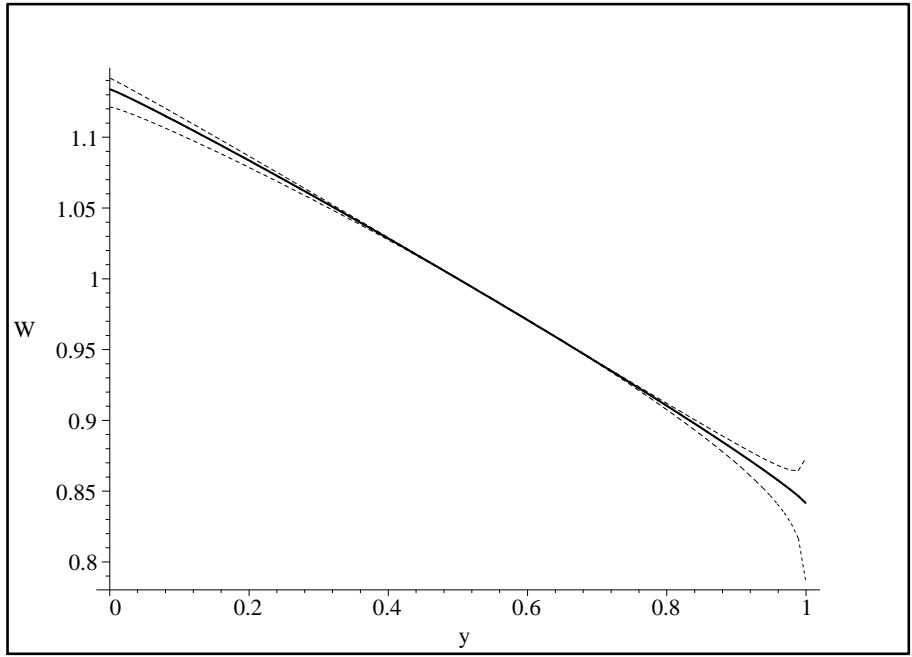}}
\caption{The solid curve corresponds to $W_1(y)$, as computed using
the numerical integration of the $F_i(y)$, as functions of $y$, for
$\kappa=1/4$, $k=0.1$, and $\Lambda=1$. The plot corresponds to
$\Omega^2=-0.0484$. (See the text for more details).}
\end{figure}

\vspace{2cm}

\begin{figure}
\vspace{2cm}
\centerline{\includegraphics[height=12cm,angle=-90]{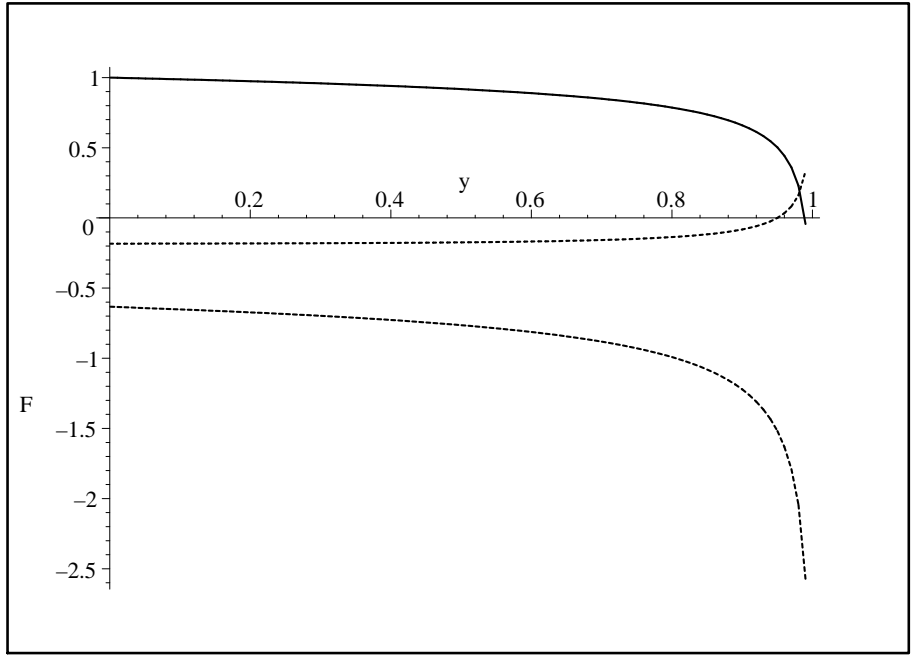}}
\caption{The curves correspond to the numerical integration of the
$F_i(y)$, as functions of $y$, for $\kappa=1/4$, $k=0.1$,
$\Lambda=1$, and $\Omega^2=-0.0484$, enforcing the boundary
conditions (\ref{num10}) at $y=0.$. (The upper curve to $F_1$,
center to $F_4$, and lower to $F_3$). Notice that the behaviour of
the $F_i$ is completely dominated by the gauge ``contamination'' as
$y \to 1 $. }
\end{figure}. \\

\vspace{2cm}

\begin{figure}
\vspace{2cm}
\centerline{\includegraphics[height=12cm,angle=-90]{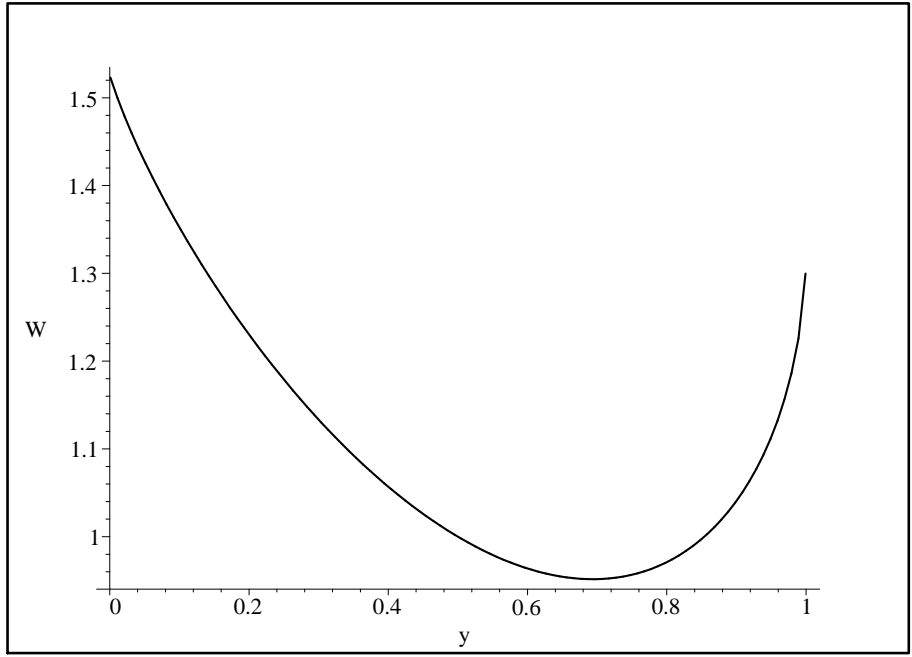}}
\caption{$W_1(y)$, as function of $y$, for $\kappa=1/4$, $k=0.1$,
$\Lambda=1$, and $\Omega^2=1.030$. }
\end{figure}

\subsection{$\kappa=1/3$.}

We also analyzed the case $\kappa=1/3$. In this case near $y=0$ we
have the expansions,
\begin{eqnarray}
\label{kappa13a}
 F_{{1}} \left( y \right)& = &a_{{0}}+{\frac {91091}{323919}}
\frac{a_0\Omega^{2}}{\Lambda}{y}^{{\frac {9}{13}}}-{\frac
{490}{4743}}a_0y+{\frac
{29575}{150784}}\frac{a_0{k}^{2}}{\Lambda}{y}^{{\frac
{16}{13}}}-{\frac
{1387350575}{19205805348}}\frac{a_0\Omega^{4}}{\Lambda^2}{y
}^{{\frac {18}{13}}}+\dots \nonumber
 \\
F_{{3}} \left( y \right)& = &-{\frac {17}{31}}a_{{0}}+{\frac
{147875}{ 323919}}\frac{{a_0\Omega}^{2}}{\Lambda}{y}^{{\frac
{9}{13}}}-{ \frac {98}{527}}a_{{0}}y+{\frac
{107653}{452352}}\frac{a_0{k}^{2}}{\Lambda}{y} ^{{\frac
{16}{13}}}-{\frac
{37843325}{446646636}}\frac{a_0{\Omega}^{4}}{\Lambda^2}{y}^{{\frac
{18}{13}}}+ \dots \nonumber
 \\
F_{{4}} \left( y \right)  &=& -{\frac {7}{31}}a_{{0}}-{\frac
{46475}{ 323919}}\frac{a_0{\Omega}^{2}}{\Lambda}{y}^{{\frac
{9}{13}}}-{ \frac {14}{1581}}a_{{0}}y-{\frac
{54925}{452352}}\frac{a_0{k}^{2}}{\Lambda}{y} ^{{\frac {16}{13}}}
\nonumber \\ & & +{\frac
{20706725}{446646636}}\frac{a_0{\Omega}^{4}}{\Lambda^2}{y}^{{\frac
{18}{13}}}+\dots
\end{eqnarray}
where dots indicate higher order terms. (The expansions were carried
out to order $y^{29/13}$ for the actual numerical integrations.)

Similarly, for $y=1$ we have,
\begin{eqnarray}
\label{kappa13b}
 F_{{1}} \left( y \right) & = & a_{{0}}+{\frac
{485199}{24272}}\,{\frac {{k
}^{2}a_{{0}}}{\Lambda}}{\eta}^{4}+{\frac {4141345}{41984}}\,{\frac
{{k }^{4}a_{{0}}}{{\Lambda}^{2}}}{\eta}^{8}+{\frac
{7138850511}{35602432}} \,{\frac
{{k}^{6}a_{{0}}}{{\Lambda}^{3}}}{\eta}^{12} \nonumber \\ && +{\frac
{ 1916151463629}{10489954304}}\,{\frac
{{k}^{8}a_{{0}}}{{\Lambda}^{4}}}{ \eta}^{16}+{\frac
{35981066372589}{1582091468800}}\,{\frac {{k}^{10}a_
{{0}}}{{\Lambda}^{5}}}{\eta}^{20}+\dots \nonumber \\
F_{{3}} \left( y \right) & = &\frac{1}{41}a_{{0}}+{\frac
{59319}{24272}}\,{ \frac {{k}^{2}a_{{0}}}{\Lambda}}{\eta}^{4}+{\frac
{14080573}{1553408}} \,{\frac
{{k}^{4}a_{{0}}}{{\Lambda}^{2}}}{\eta}^{8}-{\frac {699887305}
{35602432}}\,{\frac {{k}^{6}a_{{0}}}{{\Lambda}^{3}}}{\eta}^{12}
\nonumber \\ & &-{ \frac {77284775699703}{555967578112}}\,{\frac
{{k}^{8}a_{{0}}}{{ \Lambda}^{4}}}{\eta}^{16}-{\frac
{683640261079191}{2353843404800}}\,{ \frac
{{k}^{10}a_{{0}}}{{\Lambda}^{5}}}{\eta}^{20} +\dots \nonumber \\
F_{{4}} \left( y \right) & = &-{\frac {21}{41}}\,a_{{0}}-{\frac
{191139}{ 24272}}\,{\frac
{{k}^{2}a_{{0}}}{\Lambda}}{\eta}^{4}-{\frac {828269}{
41984}}\,{\frac {{k}^{4}a_{{0}}}{{\Lambda}^{2}}}{\eta}^{8}+{\frac {
1259797149}{35602432}}\,{\frac
{{k}^{6}a_{{0}}}{{\Lambda}^{3}}}{\eta}^ {12} \nonumber \\ &&+{\frac
{2341962899991}{10489954304}}\,{\frac {{k}^{8}a_{{0}}}{{
\Lambda}^{4}}}{\eta}^{16}+{\frac
{683640261079191}{1582091468800}}\,{ \frac
{{k}^{10}a_{{0}}}{{\Lambda}^{5}}}{\eta}^{20} +\dots
\end{eqnarray}
where $\eta=(1-y)^{1/39}$, and dots indicate higher order terms.
(For the actual computations the expansion was carried out to order
$(1-y)^{40/39}$)

We found that for $k=0.1$, and $\Lambda=1$, the lowest eigenvalue is
$\Omega^2=-0.078...$, and the next is $\Omega^2=1.22...$. The plots
of the corresponding functions $W_1$ are qualitatively similar to
those for $\kappa=1/2$, and, therefore are not shown here for
simplicity. We have therefore found that the Linet-Tian space time
for $\kappa=1/3$ is unstable, and, because of the isometry with
$\kappa=2/5$, that those space times are also unstable.

In the next section we consider the special cases $\kappa=0$, and
$\kappa=1$.

\section{The special cases $\kappa=0$ and $\kappa=1$.}

In this section we consider the particular cases $\kappa=0$, and
$\kappa=1$. Although they are isometric, since we are considering
only perturbations that do not depend on $\phi$, for $\kappa =0$, we
have a regular axis at $y=0$, while for $\kappa=1$ the axis at $y=0$
is singular and the metric is regular for $0 < y \leq 1$, i.e.,
including $y=1$.

Let us consider first $\kappa=0$. In this case (\ref{diag40})
reduces to,
\begin{equation}\label{spe02}
-{\frac {d^{2}W_1}{d{y}^{2}}} +\frac {
 \left( 3-2\,y \right)}  {
 \left( 1-y \right) y}
 {\frac {d W_1}{dy}}+{\frac {
 \left( {k}^{2}+2\,\Lambda\, \left( 1-y \right) ^{2/3} \right) }{3 y
 \left( 1-y \right) ^{5/3}\Lambda}} W_1={\frac { {\Omega}^{2}}{3y \left( 1-y \right)
 ^{5/3}\Lambda}} W_1
\end{equation}

If we introduce now a new function $\widetilde{W}_2(r)$, such that,
\begin{equation}\label{spe04}
    W_1(y)={\cal{K}}_0(y) \widetilde{W}_2(r(y))
\end{equation}
where $r(y)$ is a solution of,
\begin{equation}\label{spe06}
   \frac{dr}{dy} =\frac{1}{\sqrt{3} y^{1/2}(1-y)^{5/6}},
\end{equation}
and,
\begin{equation}\label{spe08}
    {\cal{K}}_0(y)= y^{3/4} (1-y)^{1/4},
\end{equation}
we find that if $W_1$ is a solution of (\ref{diag40}), then
$\widetilde{W}_2(r)$ is a solution of,
\begin{equation}\label{spe10}
   -\frac{d^2 \widetilde{W}_2} {dr^2}
   +{\cal{V}}_0(r)\widetilde{W}_2 = \frac{\Omega^2}{\Lambda} \widetilde{W}_2
\end{equation}
where,
\begin{equation}\label{spe12}
{\cal{V}}_0(r) = \frac{16(1-y)^{1/3} y k^2 +5 (9-8y)\Lambda}{16 y
(1-y)^{1/3} \Lambda}
\end{equation}
and it is understood that $y=y(r)$, through the inverse of
(\ref{spe06}). We notice that (\ref{spe10}) has the form of the
Schr\"odinger equation for a particle of mass $m=2$, moving in the
one dimensional potential ${\cal{V}}_0(r)$. Since ${\cal{V}}_0(r)
>0$ for $0 \leq y \leq 1$, and therefore, for the corresponding
range of $r$, then, for any acceptable boundary condition that makes
(\ref{spe10}) self - adjoint, we must have $\Omega^2 > 0$, and, as
one would expect, given that the axis $y=0$ is regular, in this case
there are no unstable modes corresponding to perturbations along
that axis, but, as we shall see, the space time is still unstable
regarding other modes. \

In the case $\kappa=1$, on the other hand, we have that the axis
$y=0$ is singular, and the equation for $W_1$ takes the form,
\begin{eqnarray}
\label{spe14} & - & {\frac {d^{2}W_1}{d{y}^{2}}} -  {\frac {
 \left( 4\,{y}^{4/3} \left( 2\,y-5 \right) {k}^{2}+ \left( 3+6\,y
 \right) \Lambda \right)  }{3
 \left( 1-y \right) y \left( 4\,{y}^{4/3}{k}^{2}-4\,\Lambda\,y+\Lambda
 \right) }}  {\frac {d W_1}{dy}}
  \nonumber \\ & + &{\frac {   \left( 4\,{y}^{5/3
}{k}^{4}+{y^{1/3}} \left( 20\,y-15-8\,{y}^{2} \right) \Lambda\,{k}^
{2}+6\, \left( 1-y \right) {\Lambda}^{2} \right) }{3 y \left( 1-y
 \right) ^{2}\Lambda\, \left( 4\,{y}^{4/3}{k}^{2}-4\,\Lambda\,y+
\Lambda \right) }} W_1 \nonumber \\ && =  {\frac { {\Omega}^{2}
}{3\Lambda\,{y}^{5/3} \left( 1-y \right) }} W_1
\end{eqnarray}

We introduce again a new function $\widetilde{W}_3(r)$, such that,
\begin{equation}\label{spe16}
    W_1(y)={\cal{K}}_1(y) \widetilde{W}_3(r(y))
\end{equation}
where now $r(y)$ is a solution of,
\begin{equation}\label{spe18}
   \frac{dr}{dy} =\frac{1}{\sqrt{3} y^{5/6}(1-y)^{1/2}},
\end{equation}
which we take as,
\begin{equation}\label{spe18a}
 r(y) = 2\sqrt{3} y^{1/6}{}_2F_1(1/6,1/2;7/6;y),
\end{equation}
where ${}_2F_1(a,b;c;{x}^{2})$ is a hypergeometric function. This
gives for $r(y)$ the (finite) range,
\begin{equation}\label{spe18b}
0 \leq r(y) \leq r_0
\end{equation}
with $r(0)=0$, and $r(1)=r_0 = 4.206...$.

The function ${\cal{K}}_1(y)$ is given by,
\begin{equation}\label{spe20}
    {\cal{K}}_1(y)= \frac{4 y^{4/3} k^2+(1-4y) \Lambda}{ y^{1/12} (1-y)^{1/4}},
\end{equation}
and we find that if $W_1$ is a solution of (\ref{diag40}), then
$\widetilde{W}_3(r)$ is a solution of the Schr\"odinger like
equation,
\begin{equation}\label{spe22}
   -\frac{d^2 \widetilde{W}_3} {dr^2}
   +{\cal{V}}_1(r)\widetilde{W}_3 = \frac{\Omega^2}{\Lambda}
   \widetilde{W}_3
\end{equation}
where,
\begin{eqnarray}\label{spe24}
{\cal{V}}_1(r) & = &\left[ 768{y}^{4}{k}^{6}+48{y}^{8/3} \left(
29-56y \right) \Lambda{k }^{4}+24{y}^{4/3} \left(68y+32\,{y}^{2}-73
\right) {\Lambda}^{2} {k}^{2} \right. \nonumber \\ & & \left.+
\left( 512{y}^{4}-528{y}^{2}-896{y}^{3}+832y-1
 \right) {\Lambda}^{3} \right] \nonumber \\
 & & \times \left[48 \left( 4{y}^{4/3}{k}^{2}+ \left(1 -4y \right) \Lambda
 \right) ^{2} \left( 1-y \right) {y^{1/3}}\Lambda \right]^{-1}
\end{eqnarray}
and it is understood that $y=y(r)$. We notice immediately that for
$k^2 < 3/4 \Lambda$ the ``potential'' ${\cal{V}}_1(r)$ has a double
pole, but, for $k^2 > 3/4 \Lambda$ it is regular for $0 < y < 1$,
and, therefore, in the corresponding range of $r$. The regular case
is important, because it allows for a self adjoint extensions of
(\ref{spe22}). To analyze this point we need the behaviour of
${\cal{V}}_1(r)$ at the boundaries $r=0$, and $r=r_0$. From
(\ref{spe18}), to leading order near $y=0$, we find,
\begin{equation}\label{spe26}
    y(r) = \frac{1}{1728} r^6 -\frac{1}{6967296}r^{12}+\dots
\end{equation}
and, therefore,
\begin{equation}\label{spe28}
   {\cal{V}}_1(r)=-\frac{1}{4 r^2}+\frac{5}{42}r^4 +\dots
\end{equation}

This implies that for the general solution of (\ref{spe22}), near
$r=0$, we would have,
\begin{equation}\label{spe28a}
 \widetilde{W}_3(r)\simeq \sqrt{r} \left[ C_1 +C_2
\ln(r)\right] ,
\end{equation}
where $C_1$, and $C_2$ are constants. This corresponds, as in
\cite{glei1} and \cite{glei2}, to the {\em circle limit} boundary
condition. For the same reasons as in \cite{glei1} and \cite{glei2},
we shall consider here only the restricted case $C_2=0$.

Using again (\ref{spe18}), near $r=r_0$, we find,
\begin{equation}\label{spe30}
    y(r) = 1-\frac{3}{4}(r_0-r)^2+\frac{5}{16}(r_0-r)^4+\dots
\end{equation}
and, then, near $r=r_0$,
\begin{equation}\label{spe32}
   {\cal{V}}_1(r)=\frac{16 k^2-3 \Lambda}{12 \Lambda
   (r_0-r)^2}+\frac{4(k^2-3\Lambda)}{9 \Lambda} +\dots
\end{equation}
which implies that, in general, near $r=r_0$,
\begin{equation}\label{spe28b}
\widetilde{W}_3(r)\simeq \sqrt{r_0-r} \left[ C_1 (r_0-r)^{\frac{2
k}{\sqrt{3\Lambda}}}+C_2 (r_0-r)^{-\frac{2 k}{\sqrt{3\Lambda}}}
\right],
\end{equation}
and, in this case we must set $C_2=0$, to have normalizable
solutions.\\

\begin{figure}
\centerline{\includegraphics[height=12cm,angle=-90]{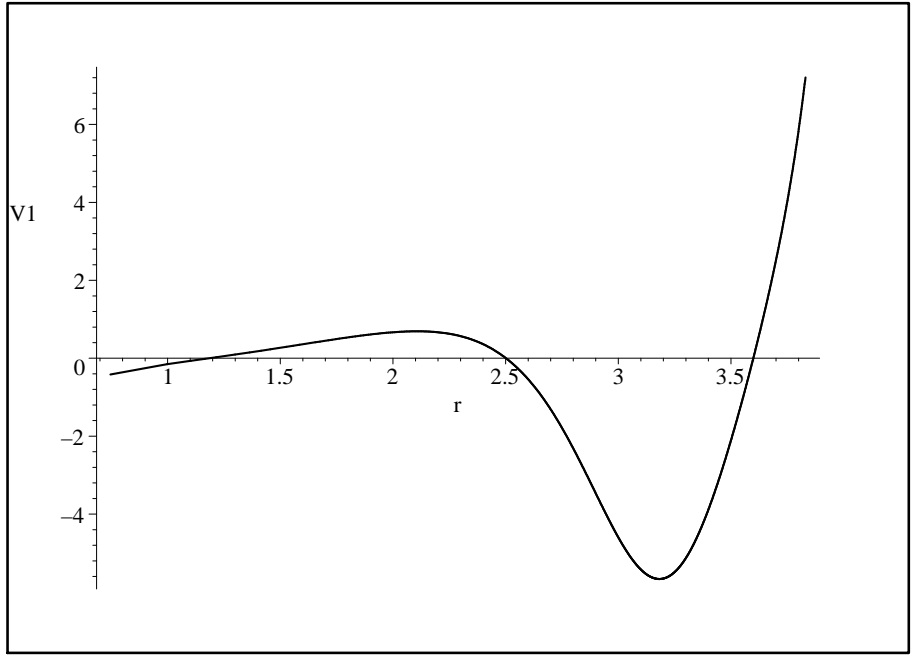}}
\caption{ ${\cal{V}}_1(r)$, as function of $r$. Notice the negative
region near $r=3$ . As indicated in the text, ${\cal{V}}_1(r)$
diverges at both $r=0$, and $r=r_0$. (Not shown in the plot).}
\end{figure}

As an example, let us take $k=1$, $\Lambda=1$. Using (\ref{spe18a}),
and (\ref{spe24}), we may easily obtain a plot of ${\cal{V}}_1(r)$
as a function of $r$. This is shown in Figure 6, where we notice the
negative region near $r=3$. This strongly suggests that there should
be at least one solution with $\Omega^2 <0$. Since we do not have an
explicit expression for $ {\cal{V}}_1(r)$ as a function of $r$, to
explore this possibility it is simpler to go back to (\ref{spe14}),
find the solutions there, and then use (\ref{spe16}) to construct
the solutions $W_3(r)$. The boundary condition (\ref{spe28a}), near
$r=0$ translates to,
\begin{eqnarray}\label{spe40}
   W_1(y)& = &a_{{0}}-{\frac {a_{{0}}{\Omega}^{2}{y^{1/3}}}{3\Lambda}}+{
\frac {9{\Omega}^{4}a_{{0}}{y}^{2/3}}{4{\Lambda}^{2}}}+{\frac {a_{{0
}} \left(8{\Lambda}^{3} -3{\Omega}^{6}\right) y}{4{\Lambda}^{3}}}
 -{\frac {9a_{{0}} \left(
20{k}^{2}{\Lambda}^{3}-{ \Omega}^{8} \right)
{y}^{4/3}}{64{\Lambda}^{4}}} \nonumber \\
 &&+{
\frac {3 a_{{0}}{\Omega}^{2} \left(
-9\,{\Omega}^{8}-960\,{\Omega}^{2}{
\Lambda}^{3}+1012\,{k}^{2}{\Lambda}^{3} \right)
{y}^{5/3}}{1600{\Lambda}^{ 5}}}
 +\dots
\end{eqnarray}
near $y=0$, and, near $y=1$, (\ref{spe28b}) translates to,
\begin{eqnarray}\label{spe42}
   W_1(y)& = & a_1\left( 1-y \right) ^{{\frac {k}{\sqrt {3\Lambda}}}}
    \left( 1-{\frac { \left( -2\,{k}^{2}+5\,\sqrt {3
\Lambda}k+3\,{\Omega}^{2}+6\,\Lambda \right) \sqrt {3}
 \left( 1-y \right) }{9 \left( 2\,k+\sqrt {3\Lambda} \right)
\sqrt {\Lambda}}}  \right. \nonumber \\ &&
 -{\frac {
 \left(10\sqrt {3\Lambda}{k}^{3}-4{k}^{4}+12{\Omega}^{
2}{k}^{2}+54\Lambda {k}^{2}+18 \sqrt {3}{\Lambda}^{3/2}k-9\,{
\Omega}^{4}-36 {\Omega}^{2}\Lambda \right)\left( 1-y \right) ^{2}
}{108 \left( 2\,{k}^{2 }+3\,\sqrt {3\Lambda}k+3 \Lambda \right)
\Lambda}}
 \nonumber \\ && \left.+\dots \right)
\end{eqnarray}
In these expressions $a_0$, and $a_1$ are arbitrary constants that
are eventually fixed when the solutions are normalized. Since the
coefficients in (\ref{spe14}) are now regular functions, it is
straightforward to apply a ``shooting'' procedure, using either the
boundary condition at $y=0$ or at $y=1$. We applied this procedure,
setting $k=1$, and $\Lambda=1$, looking for solutions from either
boundary, until we obtained coincidence, within a reasonable
numerical accuracy. The first result corresponds to the lowest
``eigenvalue'', $\Omega^2 =-1.073...$. The corresponding
``eigenfunction'', $W_3(r)$, is shown in Figure 7. We also computed
the first solution above the lowest, with $\Omega^2=1.075...$, and
$W_3(r)$ as shown in Figure 8.

\begin{figure}
\vspace{2cm}
\centerline{\includegraphics[height=12cm,angle=-90]{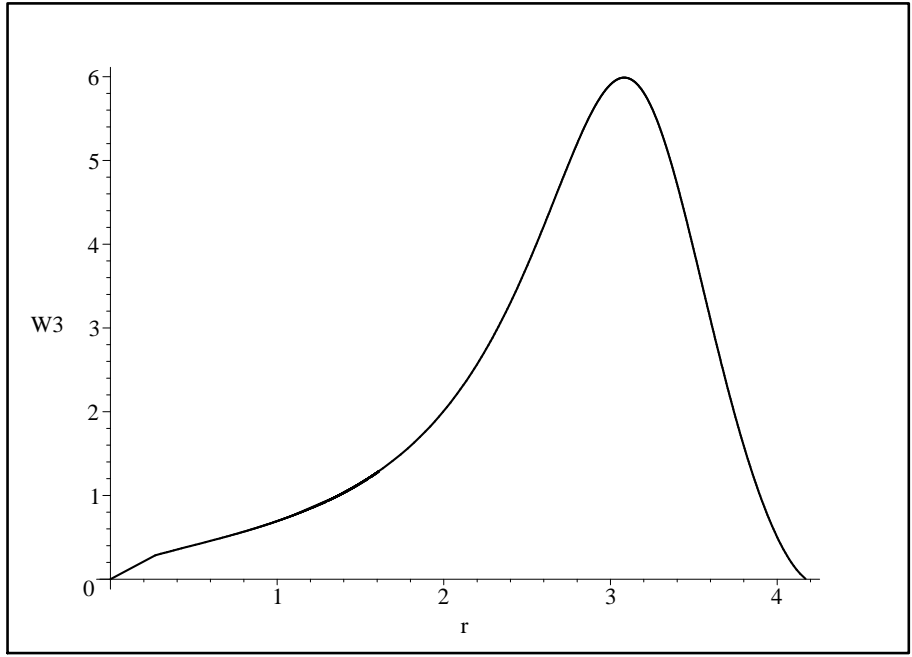}}
\caption{ $W_3(r)$ as a function of $r$, for the lowest level with
$\Lambda=1$, $k=1$, and $\Omega^2 = -1.073$. (Not normalized).}
\end{figure}

\begin{figure}
\centerline{\includegraphics[height=12cm,angle=-90]{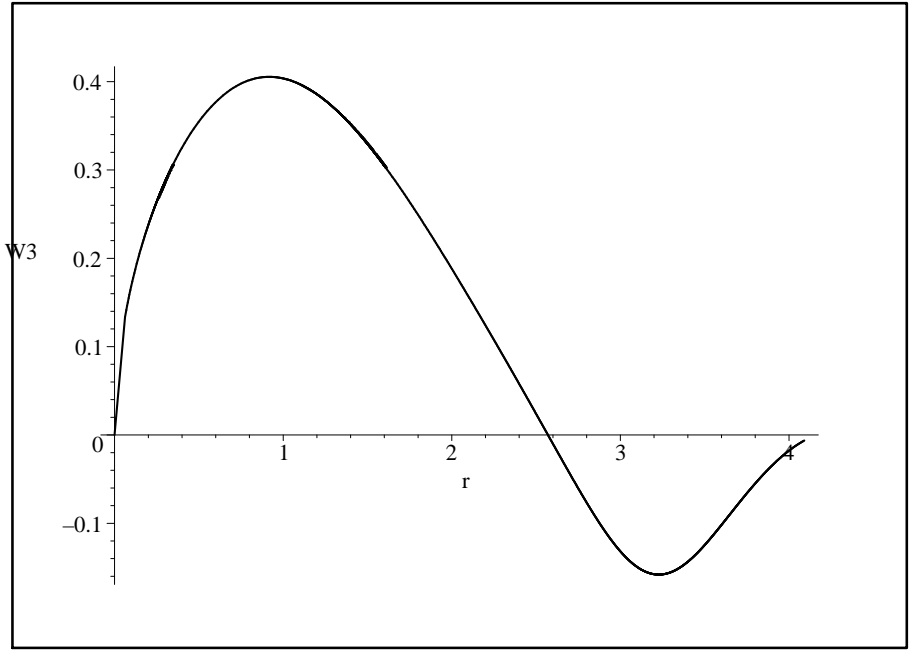}}
\caption{ $W_3(r)$ as a function of $r$, for the first level above
the lowest, with $\Lambda=1$, $k=1$, and $\Omega^2 = 1.075$. (Not
normalized).}
\end{figure}


\section{Final comments.}

In this paper we considered the Linet-Tian metrics with a positive
cosmological constant, with the purpose of extending the linear
stability analysis of \cite{glei2} to the case $\Lambda>0$. An
important difference with the case of $\Lambda <0$ is in the
structure of the resulting space time, since in the present case we
have a remarkable toroidal type symmetry, that has as a result an
isometry between metrics, with the same $\Lambda$, but where if the
other parameter is $\kappa$ in one, then it is $(1-\kappa)/(2
\kappa+1)$ in the other, with the roles of the Killing vectors
$\partial_{\phi}$, and $\partial_z$ interchanged.

For the stability analysis we introduced a new form of the metric,
and, after defining the form of the perturbations to be studied, we
gave a detailed description of their gauge dependence and related
ambiguities. The analysis of the perturbations was restricted to
what we call the ``diagonal'' case. This is characterized by four
functions, ($F_1,\; F_2,\; F_3, \;F_4$), that satisfy the linearized
Einstein equations on the background of the Linet-Tian metric. These
equations can be reduced to a set of three linear first order
ordinary differential equations for $F_1$, $F_3$, and $F_4$, but the
system is not free of gauge ambiguities. On this account we
introduced a gauge invariant function, $W_1$, which was shown to be
also a ``master function'', in terms of which one could express all
the diagonal metric perturbations. This function satisfies a linear
second order ODE, which is also linear in $\Omega^2$, where $\Omega$
is the frequency of perturbation modes, and $\Omega^2 <0$ indicates
an unstable mode. These modes are specified by imposing appropriate
boundary conditions which transform this equation in an eigenvalue -
eigenfunction problem. Unfortunately, although one can show that all
solutions $W_1$ are regular in $0 < y <1$, the coefficients of the
equation contain a singular point in that interval, where they are
divergent. As a result, it was not possible to put the equation in a
self adjoint form that would have provided with a lower bound on the
spectrum of $\Omega^2$, and an explicit form for the solution of the
initial value problem. Nevertheless, by numerically solving the
system of equations for $F_1$, $F_3$, and $F_4$, after imposing
appropriate boundary conditions at either $y=0$, or $y=1$, we could
obtain values for $\Omega^2$ that show the existence of unstable
modes for the particular values analyzed. Since the solutions, and
therefore $\Omega^2$ should depend continuously on the parameters of
the background metric, these results strongly suggest that there
should be unstable modes for the whole range $0\leq \kappa \leq 1$,
and, therefore, that all Linet - Tian space times with $\kappa$ in
the range $0\leq \kappa \leq 1$, are linearly unstable.  The problem
of determining the time evolution of arbitrary initial data in terms
of the $W_1$, or something equivalent, remains open, but we expect
to be able to solve it along the lines of \cite{dotti}. This will be
considered elsewhere.

\section*{Acknowledgments}

This work was supported in part by CONICET (Argentina). I am
grateful to G. Dotti, and A. Reula for helpful comments, suggestions
and criticisms.

 \end{document}